\documentclass{jfm}

\usepackage{graphicx}
\usepackage{epstopdf}
\usepackage{natbib}
\usepackage{amsmath,amssymb,stmaryrd,array}
\usepackage{color,soul,rotating,overpic}

\usepackage[section]{placeins}
\usepackage{array}
\usepackage{caption}
\usepackage{subcaption}
\usepackage{wrapfig}
\usepackage{color}

\ifCUPmtlplainloaded \else
  \checkfont{eurm10}
  \iffontfound
    \IfFileExists{upmath.sty}
      {\typeout{^^JFound AMS Euler Roman fonts on the system,
                   using the 'upmath' package.^^J}%
       \usepackage{upmath}}
      {\typeout{^^JFound AMS Euler Roman fonts on the system, but you
                   dont seem to have the}%
       \typeout{'upmath' package installed. JFM.cls can take advantage
                 of these fonts,^^Jif you use 'upmath' package.^^J}%
      }
  \else
  \fi
\fi

\ifCUPmtlplainloaded \else
  \checkfont{msam10}
  \iffontfound
    \IfFileExists{amssymb.sty}
      {\typeout{^^JFound AMS Symbol fonts on the system, using the
                'amssymb' package.^^J}%
       \usepackage{amssymb}%
       \let\le=\leqslant  
         
      }{}
  \fi
\fi

\ifCUPmtlplainloaded \else
  \IfFileExists{amsbsy.sty}
    {\typeout{^^JFound the 'amsbsy' package on the system, using it.^^J}%
     \usepackage{amsbsy}}
    {\providecommand\boldsymbol[1]{\mbox{\boldmath $##1$}}}
\fi

\title[Compressible flow around a circular cylinder]
{Two-dimensional compressible viscous flow around a circular cylinder}

\author[D. Canuto and K. Taira]%
{Daniel Canuto$^1$\thanks{Present address: Department of Mechanical and Aerospace Engineering, University of California, Los Angeles, CA 90095, USA}
 and 
Kunihiko Taira$^1$\thanks{Email address for correspondence: ktaira@fsu.edu}
  }

\affiliation{$^1$Department of Mechanical Engineering\\
Florida A\&M/Florida State University, Tallahassee, FL 32310, USA}

\pubyear{2013}
\volume{???}
\pagerange{???--???}
\date{?; revised ?; accepted ?. - To be entered by editorial office}
\begin{document}

\maketitle

%%%%%%%%%%%%%%%%%%%%%%%%%%%%%%%%%%%%%%%%%%%%%%%%%%%%%%%%%%%%%%%%%%

%All papers should feature a single-paragraph abstract of no more than 250 words, which provides a summary of the main aims and results. 
\begin{abstract}
Direct numerical simulation is performed to study compressible, viscous flow around a circular cylinder. The present study considers two-dimensional, shock-free continuum flow by varying the Reynolds number between 20 and 100 and the freestream Mach number between 0 and 0.5. The results indicate that compressibility effects elongate the near wake for cases above and below the critical Reynolds number for two-dimensional flow under shedding. The wake elongation becomes more pronounced as the Reynolds number approaches this critical value. Moreover, we determine the growth rate and frequency of linear instability for cases above the critical Reynolds number. From the analysis, it is observed that the frequency of the B\'enard-von K\'arm\'an vortex street in the time-periodic, fully-saturated flow increases from the dominant unstable frequency found from the linear stability analysis as the Reynolds number increases from its critical value, even for the low range of Reynolds numbers considered.  We also notice that the compressibility effects reduce the growth rate and dominant frequency in the linear growth stage. Semi-empirical functional relationships for the growth rate and the dominant frequency in linearized flow around the cylinder in terms of the Reynolds number and freestream Mach number are presented.
\end{abstract}

%%%%%%%%%%%%%%%%%%%%%%%%%%%%%%%%%%%%%%%%%%%%%%%%%%%%%%%%%%%%%%%%%%

\begin{keywords}
compressible flows, instability, wakes
\end{keywords}

%%%%%%%%%%%%%%%%%%%%%%%%%%%%%%%%%%%%%%%%%%%%%%%%%%%%%%%%%%%%%%%%%%

\section{Introduction}
\label{sec:intro}

Fluid flow around bluff bodies is one of the important flows that has been extensively studied in the field of fluid mechanics. As a basic model for such flows, and with a wide range of engineering applications in its own right, the flow over a circular cylinder has garnered much interest. The beginning of this deluge of research can be traced back to the late nineteenth and early twentieth centuries, and is owed to researchers such as \cite{Strouhal:APC78}, \cite{Rayleigh:PM79}, \cite{Benard:CAS08}, and \cite{Karman:GN11}.

It is well-known that incompressible flow over a circular cylinder can be characterized by the freestream Reynolds number $Re \equiv \rho_{\infty} u_{\infty} d / \mu_{\infty}$, where $\rho_{\infty}$ is the freestream density, $u_{\infty}$ is the freestream velocity, $d$ is the cylinder diameter, and $\mu_{\infty}$ is the freestream dynamic viscosity.  For a range of $Re$ from around $5$ to $47$, the separated flow behind the cylinder forms a steady and symmetric closed wake. The variation in wake geometry with $Re$ in this regime has been investigated in great depth for incompressible flow by \cite{Taneda:JPSJ56}, \cite{Coutanceau:JFM77a}, and \cite{Fornberg:JFM80}, among others. 

As $Re$ is increased beyond around $47$, the flow exhibits two-dimensional periodic laminar shedding of wake vortices. This regime persists up to the onset of spanwise instabilities at $Re \approx 190$ \citep{Williamson:PF88b, Behara:PF10}, and is characterized by the alternate shedding of two-dimensional vortices from the top and bottom halves of the cylinder. For $Re \gtrsim 190$, spanwise variation can be observed while the shedding of the vortices remains nominally two-dimensional. This shedding phenomenon is the well-known K\'arm\'an (B\'enard-von K\'arm\'an) vortex street. Since its first experimental observation by \cite{Mallock:PRS07} and \cite{Benard:CAS08}, a large body of work has been dedicated to its characterization. For instance, the practical importance of quantifying the unsteady loads generated by the vortex street has motivated unsteady lift and drag force measurements over a range of $Re$ from the onset of two-dimensional oscillation ($Re \approx 47$) up to and past the point at which the boundary layer becomes fully turbulent ($Re \approx 2 \times 10^6$) \citep{Fung:JASS60, Gerrard:JFM61, Bishop:PRS64, Sanada:JWE92, Norberg:JFS01}. Here, it is worth noting that the mean drag force has also been measured over a similar range of $Re$ \citep{Wieselsberger:PZ21, Finn:JAP53, Delany:NACA53, Tritton:JFM59, Roshko:JFM61} as a consequence of its significance in aerodynamics. From these measurements, it has been observed that the mean drag coefficient per unit span $C_d \equiv F_d / \frac{1}{2} \rho_{\infty} u_{\infty}^2 d$ (where $F_d$ is the mean drag force and $d$ is the cylinder diameter) remains relatively constant for $10^3 \lesssim Re \lesssim 10^5$, and that a drastic drop in $C_D$ (known as the `drag crisis') occurs at $Re \approx 2 \times 10^5$ as a consequence of delayed separation caused by the boundary layer transition to turbulence.

In addition to force measurements, a number of studies have also focused on the influence of $Re$ on the shedding frequency. A partial listing of some of the important work performed in this vein includes \cite{Roshko:NACA1191}, \cite{Tritton:JFM59}, and \cite{Williamson:PF88}. Besides determining the vortex shedding frequency, past research has also been concerned with the analysis of its stability. Such an analysis was first performed by \cite{Karman:GN11} for point vortices, and studies of a similar nature were performed for finite-cored vortices by \cite{Saffman:JFM82} and \cite{Meiron:JFM84}. These works were concerned with the determination of stable configurations for developed vortex streets, and were not intended to uncover a mechanism for the emergence of instability. \cite{Marsden:AMS76} provided the first suggestion of such a mechanism, explaining that the wake transitioning from flow with steady recirculation to that with periodic two-dimensional shedding could be understood as a Hopf bifurcation from a stable fixed point to a stable closed orbit. \cite{Provansal:JFM87} expanded on this idea, modeling the flow as a dynamical system described by the Stuart--Landau equation, given by
\begin{equation}
  \frac{\mathrm{d}A}{\mathrm{d}t} = \sigma A - \frac{1}{2}l|A|^{2}A,
  \label{eqn:SL}
\end{equation}
where $A$ is a complex-valued function of time, $\sigma = \sigma_{r} + \textrm{i}\sigma_{i}$ is a global constant of the flow that corresponds to an eigenvalue of the linearized Navier--Stokes operator, and $l = l_{1} + \textrm{i}l_{2}$ is a spatially-varying parameter related to the deformation of the mode in space as it is amplified in time (\citealp{Dusek:JFM94}). Under this model, $A$ can be any flow variable that undergoes oscillation, and is often taken as the transverse velocity $u_y$ in the literature. For a given mode, $\sigma_{r} > 0$ is the necessary condition for infinitesimal disturbances to be amplified.  This condition also corresponds to $Re > Re_c$, for which the flow is referred to as the unstable regime in this study.  In recent times, variations of the Stuart--Landau equation have been used by \cite{Noack:JFM03}, \cite{Sipp:JFM07}, and \cite{Sengupta:JFM10} to model the spatial and temporal evolution of flow instabilities. Other modern advancements in stability analysis of two-dimensional incompressible flow around a circular cylinder include the experimental determination of $\sigma_r$ \citep{Strykowski:JFM90}, the location of both the instability core and regions of high sensitivity to perturbations \citep{Giannetti:JFM07}, and the isolation of the antisymmetric modes that give rise to further instabilities at higher $Re$ \citep{Kumar:JFM09}. As can be seen from the above discussion, the wake geometry at stable $Re$, the transient behaviour of the flow as it transitions from a steady closed wake to a two-dimensional vortex street, and the fully-developed laminar vortex street have been investigated quite extensively for incompressible flow.

Considering the significant accumulation of knowledge on incompressible flow around a circular cylinder, comparatively little attention has been paid to the effects of compressibility on the behavior of the wake in these low $Re$ flow regimes.  While a limited number of studies have examined the effects of compressibility on the flow around a circular cylinder, these studies were undertaken at $Re$ on the order of $10^5$, four orders of magnitude higher than $Re_c$ \citep{Lindsey:NACA619, Macha:JOA77, Zdravkovich:1997}.  This lack of fundamental studies in low-$Re$ compressible flows may have been due to a historical dearth of engineering applications in such regimes.  However, fields are now emerging in which compressible, low-speed flows play an integral role.  For example, the design of small-scale aircraft for low-density environments, such as those seen on Mars or at ultra-high-altitudes on Earth, requires an understanding of such flows.  Basic aerodynamic information in these areas is currently limited but is starting to be compiled \citep{Drela:JOA92, Okamoto:Thesis05, Suwa:AIAA12, Nagai:AIAA13, Munday:JA15}. Another example is the optimization of liquid atomization, where micro-sized droplets travelling at relatively high speeds are formed. While the overall flow is a complex multiphase problem \citep{Gorokhovski:ARFM08, Shinjo:IJMF10}, the behaviour of these droplets and other small structures is partially a consequence of compressibility effects at low $Re$. It is thus evident that in order to expand the horizon of fluid mechanics applications, a fundamental knowledge of low-$Re$ compressible flow is necessary.

The present study employs two-dimensional direct numerical simulation (DNS) in an attempt to fill this gap in our knowledge of the effects of compressibility on the two-dimensional, viscous flow around a circular cylinder at Reynolds numbers near the onset of two-dimensional shedding. In particular, we aim to uncover these effects with respect to the characteristic parameters of steady wakes, the dominant frequency of shedding wakes, and the primary wake instability that causes a transition between the two wake types for $Re > Re_c \approx 47$.  Such knowledge will be useful not only for gaining a deeper comprehension of this fundamental flow, but also for validation of compressible flow solvers and stability analysis.

In what follows, we describe the problem of interest and discuss the findings from the present study. The computational setup, numerical method, and validation are discussed in \S \ref{sec:method}. The results based on the computation are presented in \S \ref{sec:results}, where we illustrate the influence of compressibility on the wake behind a circular cylinder. Findings from compressible flow near the onset of shedding are presented.  Also discussed in detail are the characteristics of the shedding instability that appears for $Re>Re_c$ and the effects of compressibility on these properties. To complement this discussion, semi-empirical relationships for the growth rate and dominant unstable frequency at the onset of this instability in terms of $Re$ and $M_{\infty}$ are presented. Concluding remarks are offered in \S \ref{sec:conclusion}.

%%%%%%%%%%%%%%%%%%%%%%%%%%%%%%%%%%%%%%%%%%%%%%%%%%%%%%%%%%%%%%%%%%

\section{Simulation approach}
\label{sec:method}

\subsection{Problem description}

The present investigation numerically examines the two-dimensional, compressible, viscous flow around a circular cylinder. The governing equations for this flow are the full compressible Navier-Stokes equations:
\begin{gather}
\frac{\partial \hat{\rho}}{\partial \hat{t}} +  \frac{\partial }{\partial \hat{x}_j} (\hat{\rho} \hat{u}_j) =  0, \\
\frac{\partial(\hat{\rho} \hat{u}_i)}{\partial \hat{t}} 
+ \frac{\partial }{\partial \hat{x}_j} (\hat{\rho} \hat{u}_i \hat{u}_j + \hat{p}\delta_{ij}) 
=  \frac{1}{Re_{a_{\infty}}} \frac{\partial}{\partial \hat{x}_j} 
\left(\frac{\partial \hat{u}_i}{\partial \hat{x}_j} +\frac{\partial \hat{u}_j}{\partial \hat{x}_i} 
- \frac{2}{3}\frac{\partial \hat{u}_k}{\partial \hat{x}_k}\delta_{ij}\right),
\end{gather}
\begin{equation}
\begin{split}
\frac{\partial \hat{e}}{\partial \hat{t}} 
& + \frac{\partial }{\partial \hat{x}_j} \left[ (\hat{e}+\hat{p})\hat{u}_j \right] \\
& = \frac{1}{Re_{a_{\infty}}} \frac{\partial}{\partial \hat{x}_j} 
          \left[\hat{u}_i\left(\frac{\partial \hat{u}_i}{\partial \hat{x}_j} 
          +\frac{\partial \hat{u}_j}{\partial \hat{x}_i} 
          - \frac{2}{3}\frac{\partial \hat{u}_k}{\partial \hat{x}_k}\delta_{ij}\right)\right] 
         + \frac{1}{Re_{a_{\infty}}}\frac{1}{Pr}\frac{\partial^2 \hat{T}}{\partial \hat{x}_k \partial \hat{x}_k},
\end{split}
\end{equation}
where variables with a hat (e.g., $\hat{\rho}$) have been non-dimensionalized according to
\begin{equation*}
  \hat{\rho} = \frac{\rho}{\rho_{\infty}}, \quad \hat{p} = \frac{p}{\rho_{\infty}a_{\infty}^2}, \quad \hat{T} = \frac{T}{T_{\infty}}, \quad \hat{e} = \frac{e}{\rho_{\infty}a_{\infty}^2},
\end{equation*}
\begin{equation}
  \hat{u}_i = \frac{u_i}{a_{\infty}}, \quad \hat{x}_i = \frac{x_i}{d}, \quad \hat{t} = \frac{ta_{\infty}}{d}.
\end{equation}
Here, $a_\infty$ is the freestream sonic speed, $\rho_\infty$ is the freestream density, $T_{\infty}$ is the freestream temperature, and $d$ is the cylinder diameter. Note that the origin of the spatial coordinate system is placed at the center of the cylinder. The dimensionless parameters that appear in the governing equations are the acoustic Reynolds number and Prandtl number, given by
\begin{equation}
  Re_{a_{\infty}} \equiv \frac{\rho_{\infty} a_{\infty} d}{\mu_{\infty}} \quad \textrm{and} \quad Pr = \frac{\mu_{\infty}}{\rho_{\infty} \alpha_{\infty}},
\end{equation}
respectively, where $\mu_{\infty}$ is the freestream dynamic viscosity and $\alpha_{\infty}$ is the freestream thermal diffusivity. In this study, the flow fields are specified by the Mach number and convective Reynolds number, defined as
\begin{equation}
 M_\infty \equiv \frac{u_\infty}{a_\infty} \quad  \textrm{and} \quad Re \equiv \frac{\rho_{\infty} u_{\infty} d}{\mu_{\infty}} = Re_{a_{\infty}}M_{\infty},
\end{equation}
respectively, where $u_{\infty}$ is the freestream velocity. The ranges of Reynolds number and Mach number considered in this study are $20 \le Re \le 100$ and $0 \le M_\infty \le 0.5$, respectively, with $M_\infty = 0$ representing the incompressible limit of the flow. These values are chosen to examine the influence of Mach number on the wake and the emergence of instability. Throughout the study, we use $\gamma = 1.4$ for the specific heat ratio, representative of air.

The forces on the cylinder are reported in terms of the drag and lift coefficients per unit span
\begin{equation}
   C_d \equiv \frac{F_d}{\frac{1}{2}\rho_\infty u_\infty^2 d} 
   \quad \textrm{and} \quad
   C_l \equiv \frac{F_l}{\frac{1}{2}\rho_\infty u_\infty^2 d},
\end{equation}
where $F_d$ and $F_l$ are the sectional drag and lift forces, respectively. In the following discussions, an overbar ($\overline{\left.~\right.}$) applied to these coefficients indicates a time-averaged value, while a prime ($~'$) denotes a single-sided fluctuation amplitude. The pressure is non-dimensionalized as the coefficient of pressure using the far-field dynamic pressure
\begin{equation}
   C_p \equiv \frac{p-p_\infty}{\frac{1}{2}\rho_\infty u_\infty^2}.
\end{equation}
To analyze periodic shedding in the unstable regime after the flow has reached its limit cycle, we use Strouhal number as the non-dimensional frequency
\begin{equation}
  St \equiv \frac{fd}{u_{\infty}},
\end{equation}
where $f$ is the dimensional shedding frequency.  

The range of Knudsen number $Kn = \sqrt{\pi \gamma/2} M_{\infty}/Re$ for the cases considered in this study is $0$ to $0.037$ (with the maximum $Kn$ occurring for the case of $M_{\infty} = 0.5$ and $Re = 20$).  Since $Kn$ characterizes the ratio of the molecular mean free path to the characteristic length, the continuum assumption may be applied for $Kn \ll 1$. In the present study, we therefore treat the flow as a continuum with no-slip applied at the cylinder surface \citep{Bird94}.  While there may be some molecular effects that can contribute to the flow physics, we limit the current investigation to examine the compressibility effects on the cylinder wake and its stability properties in the context of a continuum.

%%%%%%%%%%%%%%%%%%%%%%%%%%%%%%%%%%%%%%%%%%%%%%%%%%%%%%%%%%%%%%%%%%
\subsection{Numerical method}

For the present study, we perform direct numerical simulations (DNS) using the compressible flow solver {\it CharLES}, developed by Cascade Technologies, Inc. Detailed description of the code is provided in \cite{Khalighi:AIAA11} and \cite{Bres:AIAA2012}. {\it CharLES} uses a second-order accurate finite-volume method in conjunction with third-order Runge-Kutta time integration scheme. Extensive validation using this solver has been performed against compressible flows over bodies at low-to-moderate $Re$.

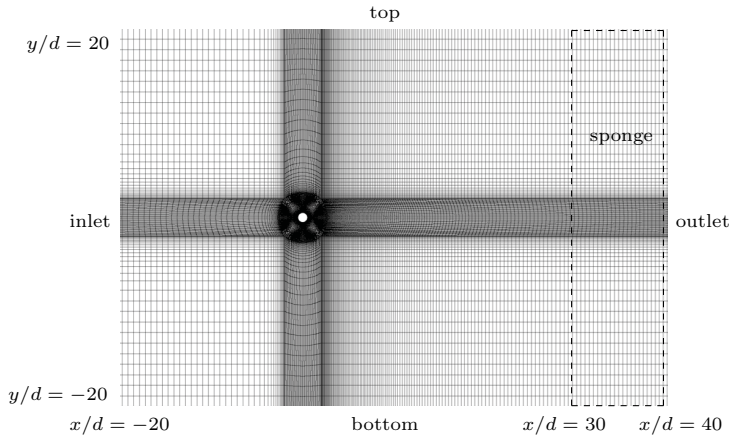
\begin{figure}
  \begin{center}
    \vspace{9pt}
    {\scriptsize
    \begin{overpic}[trim = 3cm 0.7cm 0cm 0cm, clip = true, width=0.6\textwidth]{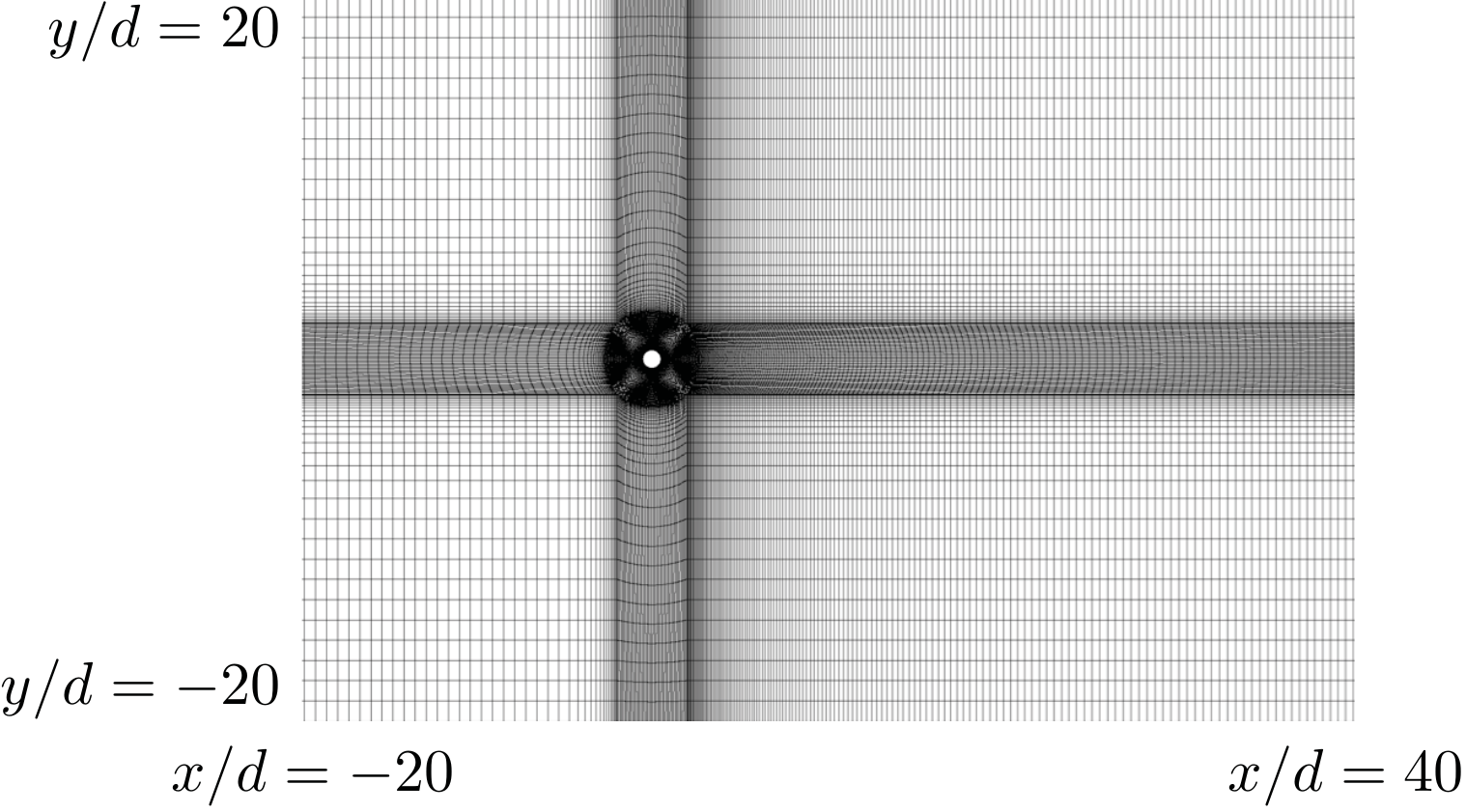} % see overpic for details.
       \put(-17,3){$y/d = -20$} % add grid to [grid, width=...] above to see coordinates
       \put(-14,60){$y/d = 20$}
       \put(-7,-2){$x/d = -20$}
       \put(67,-2){$x/d = 30$}
       \put(86,-2){$x/d = 40$}
       \put(42,65){top}
       \put(39,-2){bottom}
       \put(-7,31){inlet}
       \put(92,31){outlet}
       \put(78,45){sponge}
       \put(75,2){\dashbox{1}(15,61)}
    \end{overpic}}
    \vspace{2mm}
    \caption{Representative computational domain used for this study.}
    \label{fig:grid}
  \end{center}
\end{figure}

The computational domain in this study is chosen to be sufficiently large with a dimension of $(x/d, y/d) \in [-20,40] \times [-20,20]$, as shown in figure \ref{fig:grid}. The flow is prescribed to be uniform at the inlet $\boldsymbol{u}/u_\infty = (1, 0)$ with freestream values of non-dimensional density and pressure being specified as $1$ and $1/\gamma$, respectively. Along the outlet, a sponge zone is applied for the exiting wake vortices to be damped out without affecting the near-field solution \citep{Freund:AIAAJ97}. For the top and bottom boundaries, the symmetry boundary conditions are specified. On the cylinder surface, we prescribe the no-slip and adiabatic boundary conditions.  To ensure stability of the method, the CFL number, defined as $\textrm{CFL} = a_{\infty} \Delta t/ \Delta x_{\min}$ is limited to $1$, where $\Delta x_{\min}$ is the minimum streamwise grid spacing and $\Delta t$ is the time step.  

\begin{table}
  %\vspace{-10pt}
  \centering
  \begin {tabular}{ccccc}
    Grid size     	    				&      	$\overline{C}_d$ 	& $\Delta \overline{C}_d$ 	&	$C_{l}^{\prime}$	& $\Delta C_{l}^{\prime}$ \\ \hline % \\ [0.5ex] \hline	
    Baseline	($9.2\times 10^4$ points) 	&       $1.378$ 		& $0.073\%$ 			&	$0.325$		& $-0.612\%$ \\	
    Refined 	($1.9\times 10^6$ points) 	&       $1.377$ 		& - 					&	$0.327$		& - \\	
   \hline
  \end{tabular}
  \caption {Computed drag and lift coefficients from a grid refinement study for $Re = 100$ and $M_{\infty} = 0.25$.  The differences, $\Delta \overline{C}_d$ and $\Delta {C}_l$, are calculated as the difference from the refined case. }
  \label{table:gridref}
\end{table}

\begin{table}
  %\vspace{-10pt}
  \centering
    \begin {tabular}{ccccc}
    Grid size     	    						&      	$\overline{C}_d$ 	& $\Delta \overline{C}_d$ 	&	$C_{l}^{\prime}$	& $\Delta C_{l}^{\prime}$ \\ [0.5ex] \hline	
    Baseline	($(x/d, y/d) \in [-20,40] \times [-20,20]$) 	&       $1.350$ 		& $0.521\%$ 			&	$0.329$		& $-0.604\%$ \\	
    Large 	($(x/d, y/d) \in [-20,60] \times [-40,40]$) 	&       $1.343$ 		& -	 				&	$0.331$		& - \\	
  \hline
  \end{tabular}
  \caption {Computed drag and lift coefficients from a domain size study for $Re = 100$ and $M_{\infty} = 0.05$. The differences, $\Delta \overline{C}_d$ and $\Delta {C}_l$,  are calculated as the difference from the baseline case. }  
  \label{table:domainsize}
\end{table}

\begin{table}
  \centering
    \begin {tabular}{lllccccccc}
    $Re$			& $M_{\infty}$			& References				& $l_r/d$	& $l_a/d$	& $l_b/d$	& $\theta_s$	& 	$\overline{C}_d$ 	& $C_{l}^{\prime}$ 	& $St$ 	\\ \hline
    $20$			& $0$					& \cite{Coutanceau:JFM77a}		& $0.93$	& $0.33$	& $0.46$	& $45.0^\circ$	& - 	   		& -				& -		\\
				&					& \cite{Tritton:JFM59}			& -		& -		& -		& -			& $2.09$ 		& -				& -		\\
				&					& \cite{Linnick:JCP05}			& $0.93$	& $0.36$	& $0.43$	& $43.5^\circ$ 	& $2.06$  		& -				& -		\\
				&					& \cite{Taira:JCP07}			& $0.94$	& $0.37$	& $0.43$	& $43.3^\circ$	& $2.06$ 		& -				& -		\\
				&					& Present					& $0.92$	& $0.36$	& $0.42$	& $43.7^\circ$ 	& $2.07$		& -				& -		\\ \hline	
    $40$			& $0$					& \cite{Coutanceau:JFM77a}		& $2.13$	& $0.76$	& $0.59$	& $53.8^\circ$	& - 			& -				& -		\\
				& 					& \cite{Tritton:JFM59}			& -		& -		& -		& -			& $1.59$  		& -				& -		\\
				& 					& \cite{Linnick:JCP05}			& $2.28$	& $0.72$	& $0.60$	& $53.6^\circ$ 	& $1.52$		& -				& -		\\
				& 					& \cite{Taira:JCP07}			& $2.30$	& $0.73$	& $0.60$	& $53.7^\circ$	& $1.54$ 		& -				& -		\\
				& 					& Present					& $2.24$	& $0.72$	& $0.59$	& $53.7^\circ$ 	& $1.54$		& -				& -		\\ \hline
    $100$			& $0.05$ 				& \cite{Karagiozis:JCP10}			& -		& -		& -		& -			& $1.317$		& $0.320$			& $0.168$	\\
				&					& Present					& -		& -		& -		& -			& $1.350$		& $0.329$			& $0.167$	\\ \hline
    $100$			& $0.25$ 				& \cite{Karagiozis:JCP10}			& -		& -		& -		& -			& $1.336$		& $0.319$			& $0.168$	\\
				&					& Present					& -		& -		& -		& -			& $1.378$		& $0.325$			& $0.163$	\\ \hline
    \end{tabular}
  \caption {Comparison of characteristic wake properties against values obtained from past studies.  See Figure \ref{fig:cylinder_wake} for definitions of $l$, $a$, $b$, and $\theta_s$.}
  \label{table:comparison_literature}
\end{table}

To test the validity of our computational setup, grid resolution and domain size were examined. Owing to the availability of reference conditions, the validation conditions were chosen to be $Re = 100$ and $M_{\infty} = \{0.05, 0.25\}$. The $C_d$ and $C_l$ values obtained from the grid refinement study are shown in Table \ref{table:gridref}. From these results, it can be observed that the baseline grid provides sufficient resolution. Using the same baseline grid, the changes in $C_d$ and $C_l$ due to variation in domain size are displayed in Table \ref{table:domainsize}. These results indicate that our baseline computational domain is sufficiently large to prevent the computational boundaries from influencing the flow near the cylinder.

\begin{figure}
  \centerline{\includegraphics[width=0.475\textwidth]{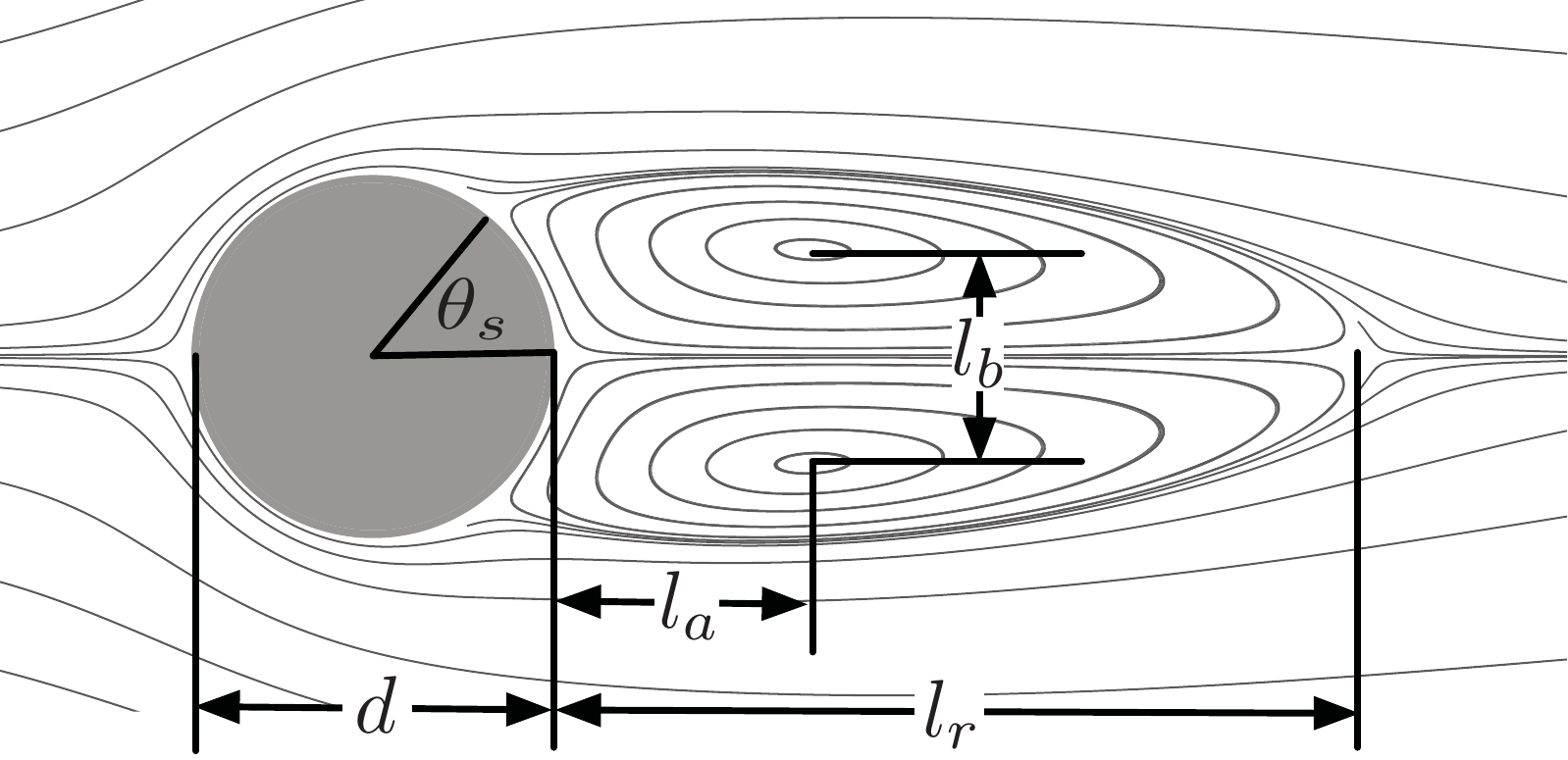}}
  \caption{Characteristic wake properties for steady flow behind a circular cylinder.}
  \label{fig:cylinder_wake}
\end{figure}

Incompressible and compressible flows were simulated on the baseline setup and compared with those reported by several other studies \citep{Coutanceau:JFM77a, Tritton:JFM59, Linnick:JCP05, Taira:JCP07, Karagiozis:JCP10}. The results of validation for different flows are shown in Table \ref{table:comparison_literature}, where the characteristic wake properties $l_r$, $l_a$, $l_b$, and $\theta_s$ are defined in figure \ref{fig:cylinder_wake}. From the comparison, it can be seen that the baseline grid and computational domain size allow for good agreement with previous work for flows both above and below the critical transition for two-dimensional shedding ($Re_c \approx 47$ for $M_{\infty} = 0$ in this study). The baseline setup is hence used for the remainder of the study.

%%%%%%%%%%%%%%%%%%%%%%%%%%%%%%%%%%%%%%%%%%%%%%%%%%%%%%%%%%%%%%%%%%

\section{Compressible flow over a circular cylinder}
\label{sec:results}
		
In this section, we first divide our attention between the flow field characteristics of stable and unstable flows. We then revisit the emergence of instability with linear stability analysis.  Stable flows are defined for $Re < Re_c$ and the flow is steady.  On the other hand, for unstable flows, we have $Re > Re_c$ and the flow exhibits unsteady vortex shedding. In our study, $Re_c \approx 47$ for the incompressible limit, though the following discussion will show that this value exhibits a slight sensitivity to $M_{\infty}$.

%%%%%%%%%%%%%%%%%%%%%%%%%%%%%%%%%%%%%%%%%%%%%%%%%%%%%%%%%%%%%%%%%%

\subsection{Flow physics}

\begin{figure}
  \begin{center}  
  \begin{tabular}{m{0.13\textwidth}m{0.43\textwidth}m{0.43\textwidth}m{0.0001\textwidth}} \hline
  & \center{$Re = 20$} & \center{$Re = 40$} & \\ \hline
  $M_{\infty} = 0$ &
  \includegraphics[trim = 0cm 0cm 9cm 0cm, clip = true, width=0.4\textwidth]{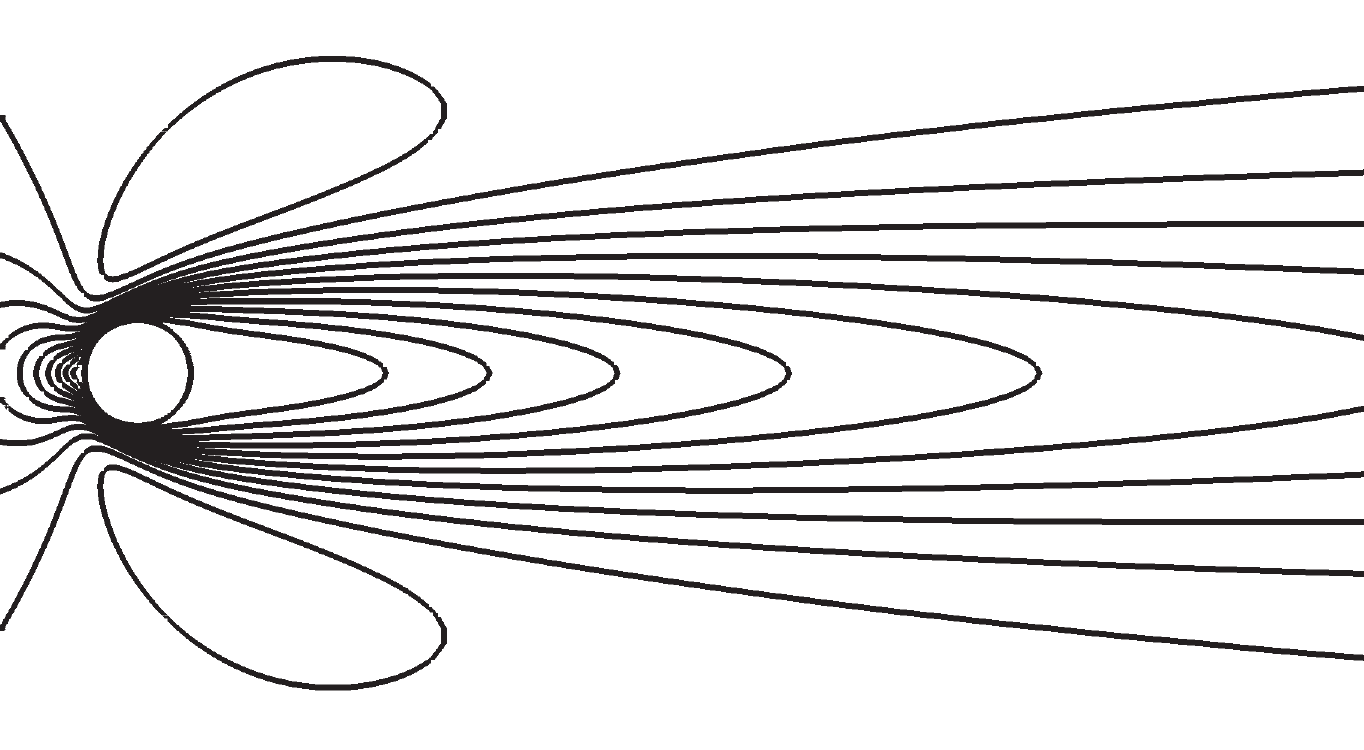} & 
  \includegraphics[trim = 0cm 0cm 9cm 0cm, clip = true, width=0.4\textwidth]{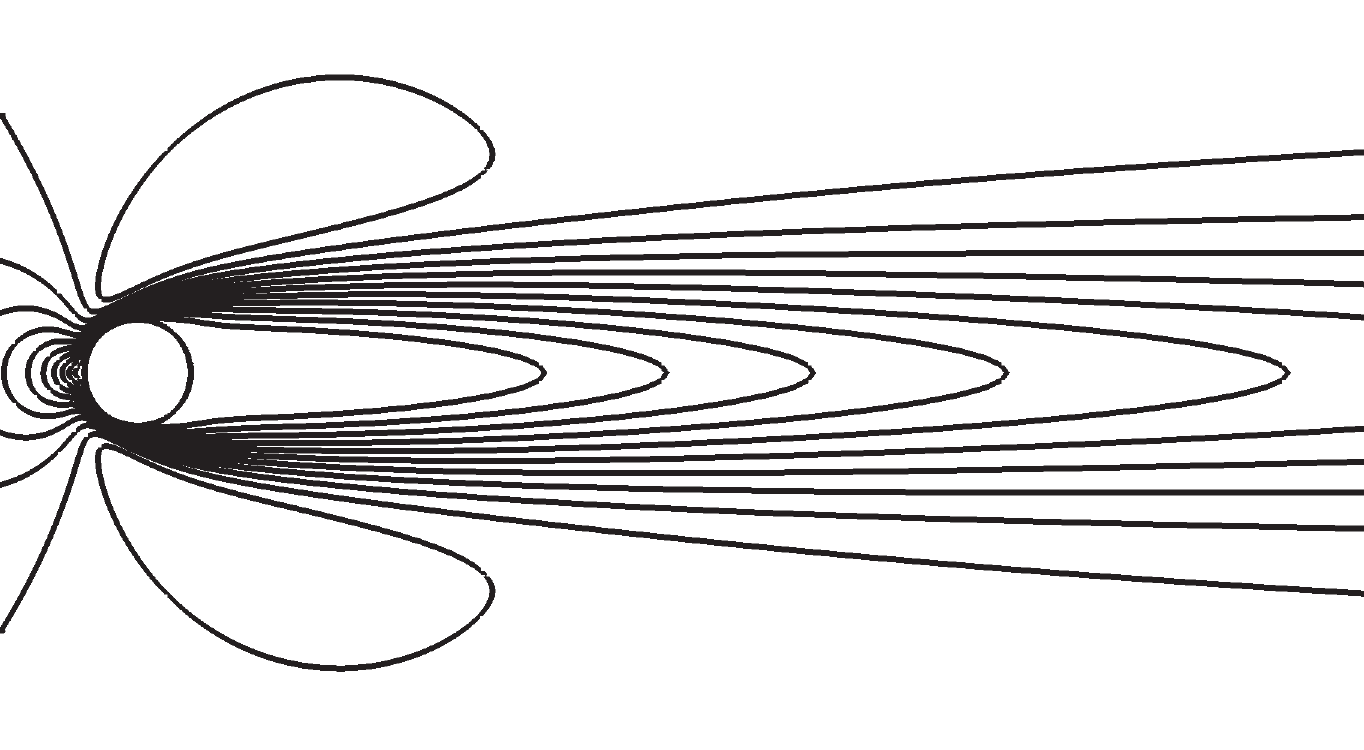} 	\\
  $M_{\infty} = 0.3$ &
  \includegraphics[trim = 0cm 0cm 9cm 0cm, clip = true, width=0.4\textwidth]{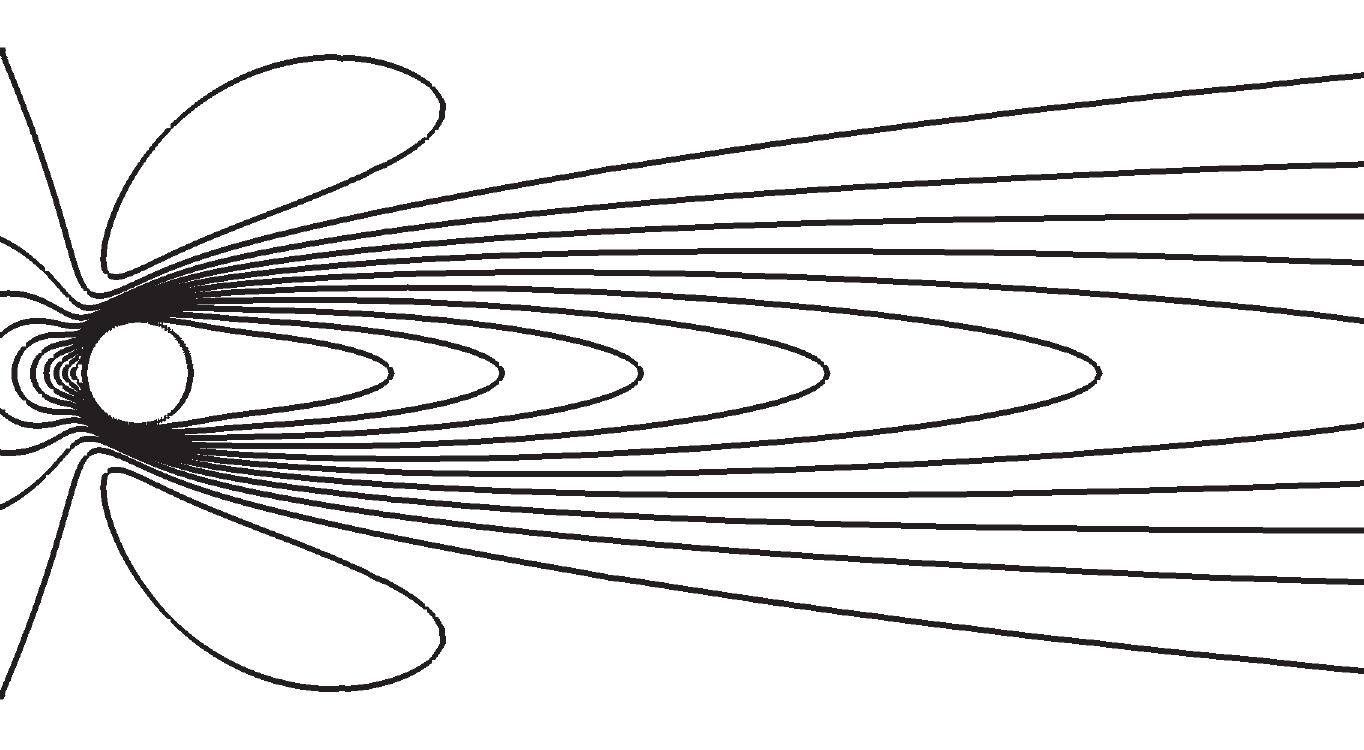} &
  \includegraphics[trim = 0cm 0cm 9cm 0cm, clip = true, width=0.4\textwidth]{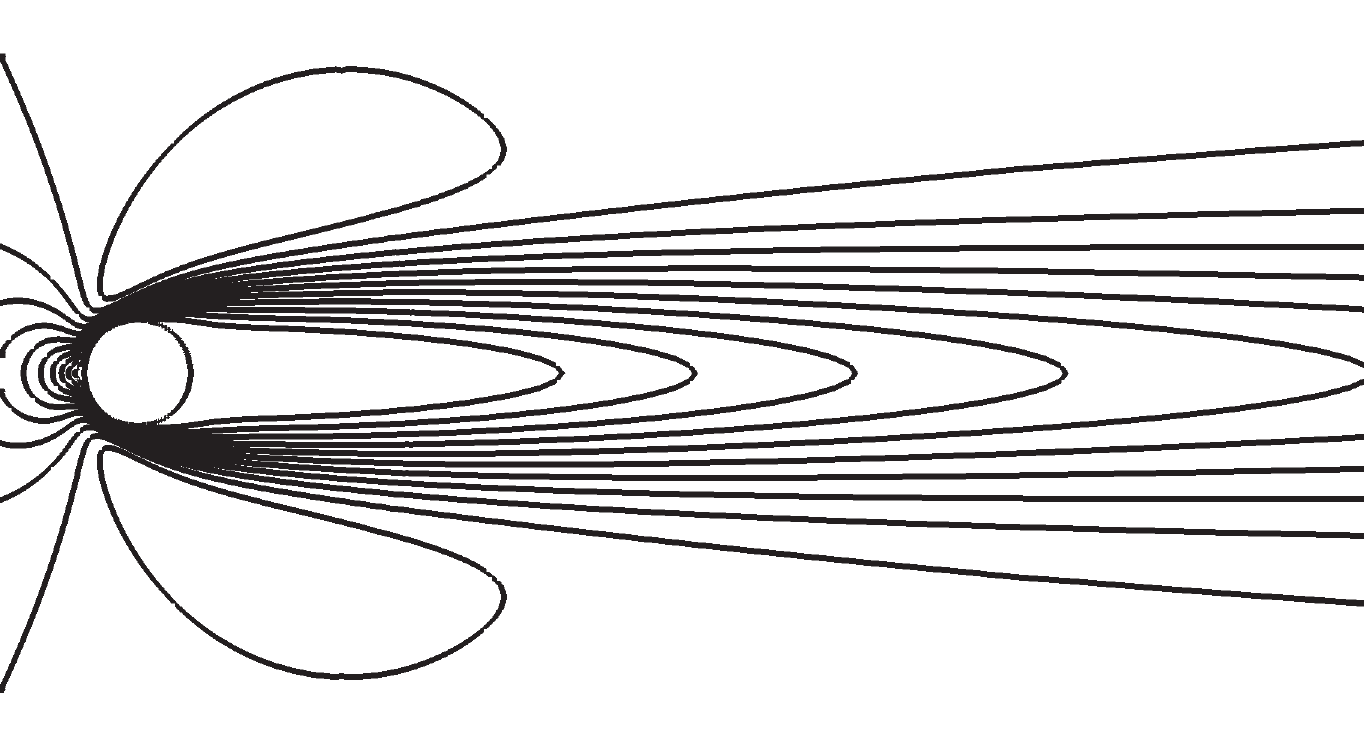} \\
  $M_{\infty} = 0.5$ &
  \includegraphics[trim = 0cm 0cm 9cm 0cm, clip = true, width=0.4\textwidth]{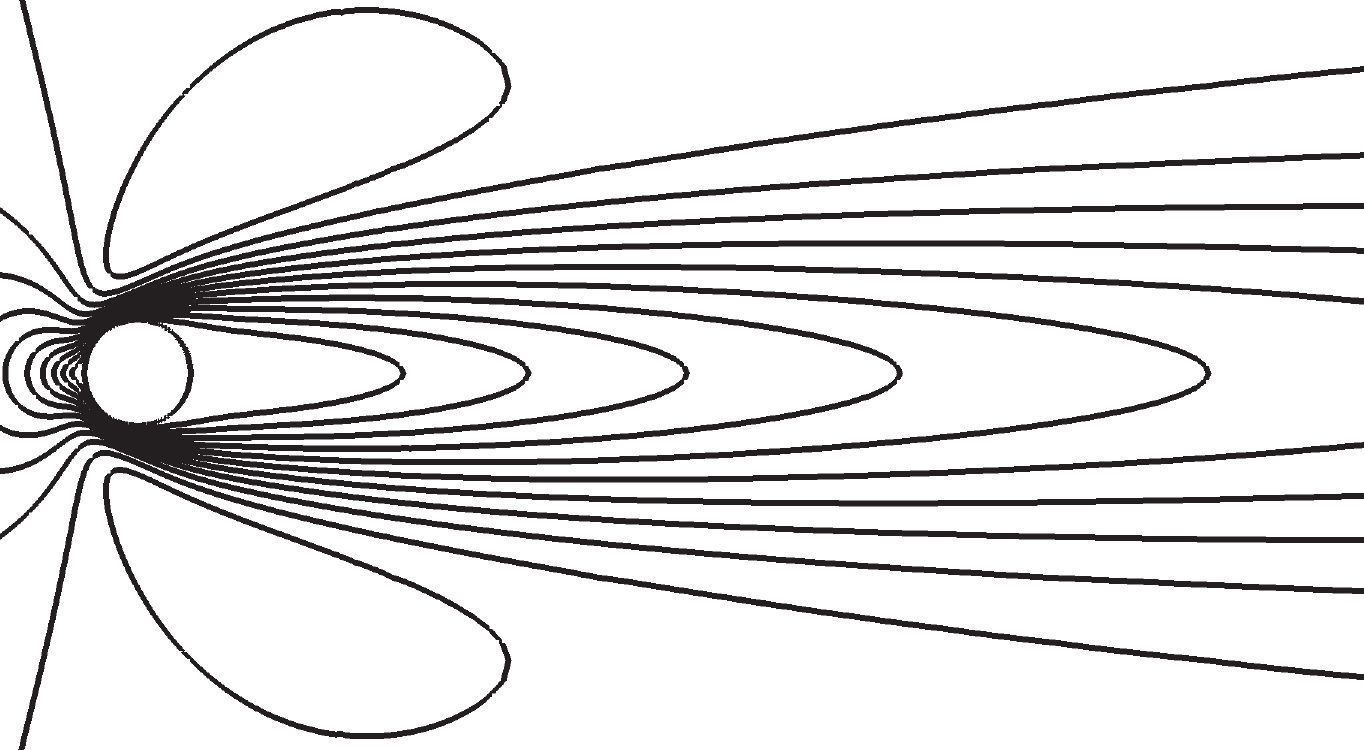} &
  \includegraphics[trim = 0cm 0cm 9cm 0cm, clip = true, width=0.4\textwidth]{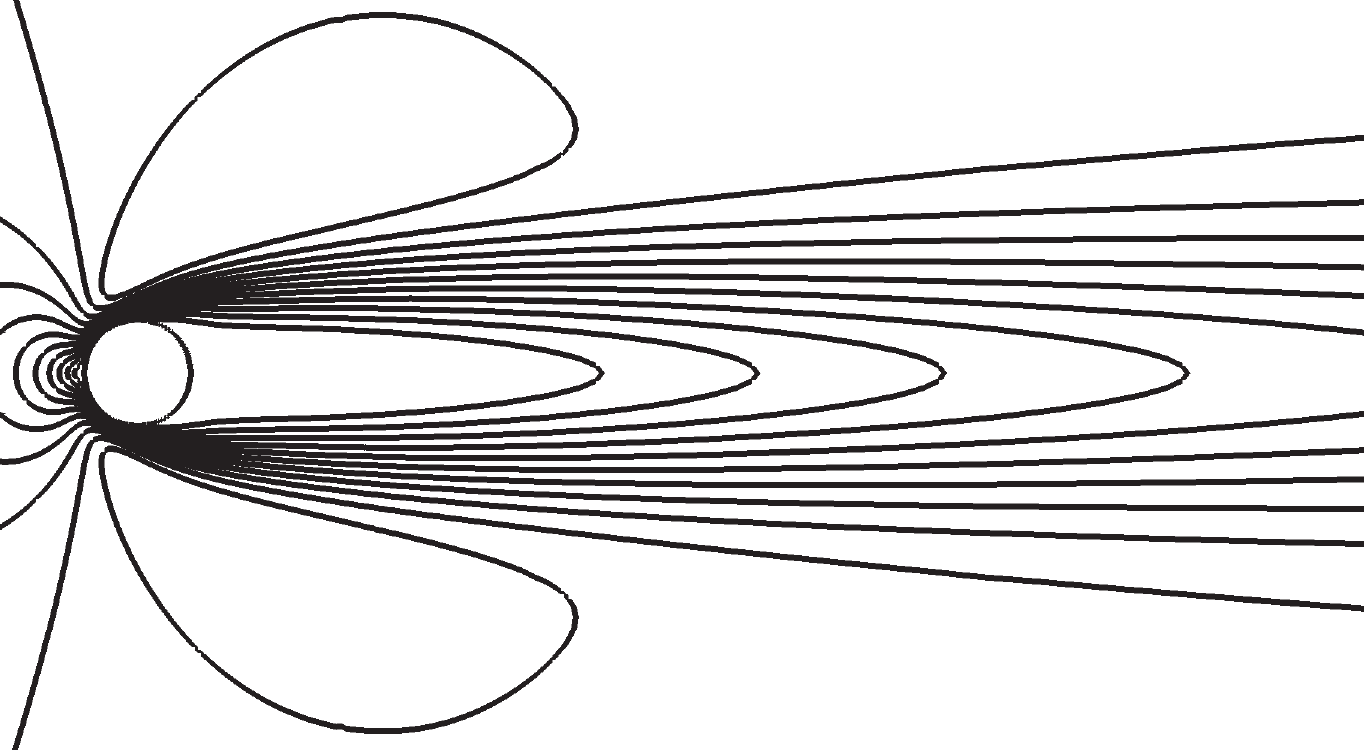} \\ \hline %\trim goes CCW from left side
  \end{tabular} 
  \caption{$\|\textbf{u}\|/u_{\infty}$ for varying $M_{\infty}$ at stable $Re$. Contours are from 0 to 1.1 across 12 levels, and flow is oriented from left to right.} 
  \label{fig:steady_vbarmag}
  \end{center}
\end{figure}

\begin{figure}
  \begin{center}
  \begin{tabular}{m{0.009\textwidth}m{0.48\textwidth}m{0.009\textwidth}m{0.48\textwidth}m{0.0001\textwidth}}
    & \center{$Re = 20$} & 	& \center{$Re = 40$} & \\
    \rotatebox[origin=c]{90}{$l_a/d$} & 
    {\scriptsize
    \begin{overpic}[width=0.48\textwidth]{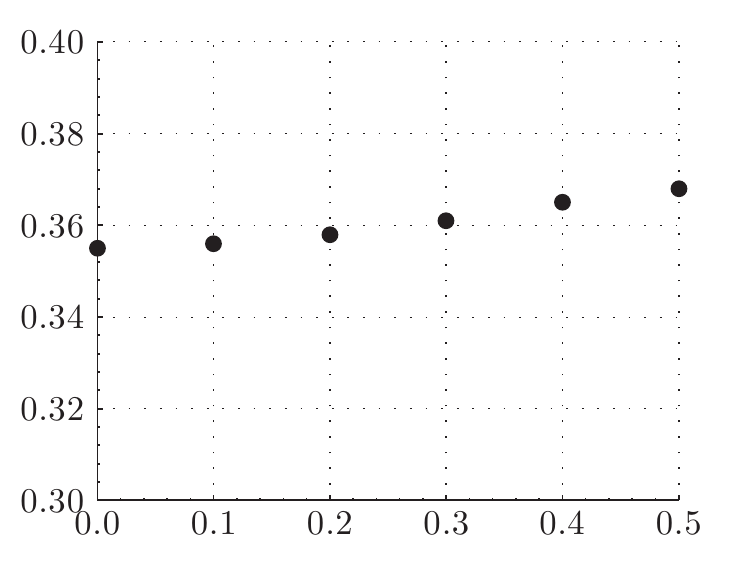} % see overpic for details.
    \end{overpic}} &	
    \rotatebox[origin=c]{90}{$l_a/d$} & 
    {\scriptsize
    \begin{overpic}[width=0.48\textwidth]{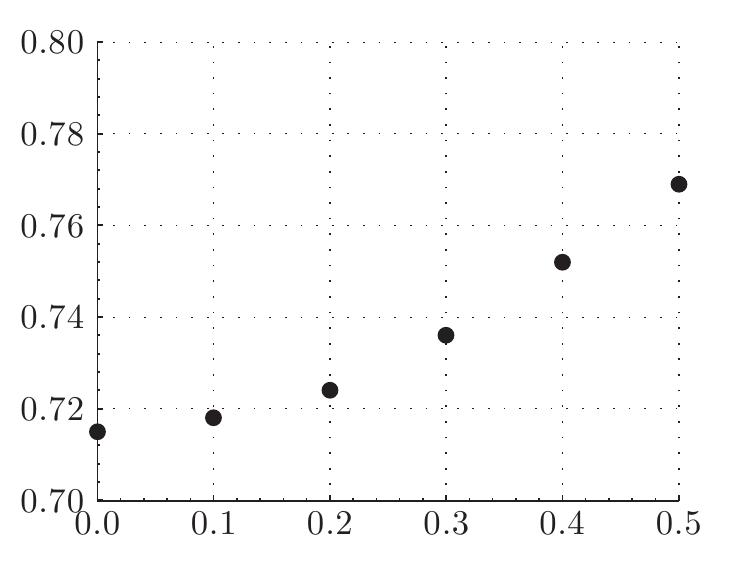} % see overpic for details.
    \end{overpic}} \\
    \rotatebox[origin=c]{90}{$l_r/d$} & 
    {\scriptsize
    \begin{overpic}[width=0.48\textwidth]{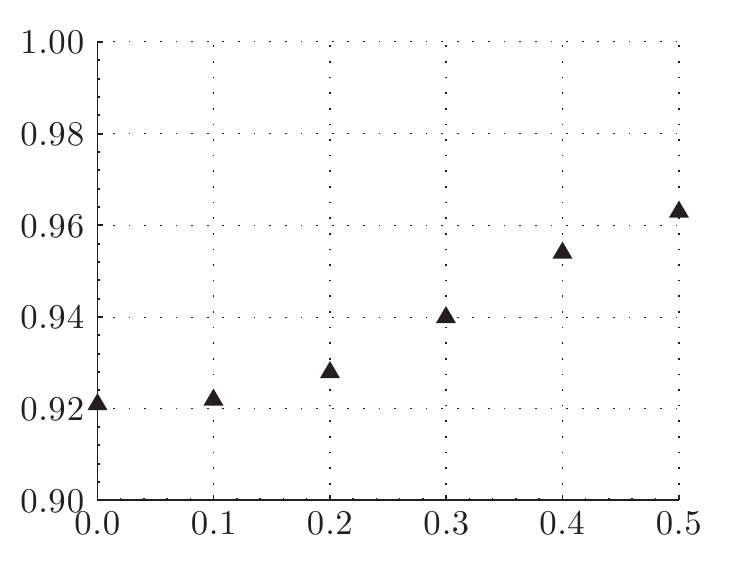} % see overpic for details.
    \end{overpic}} &	
    \rotatebox[origin=c]{90}{$l_r/d$} & 
    {\scriptsize
    \begin{overpic}[width=0.48\textwidth]{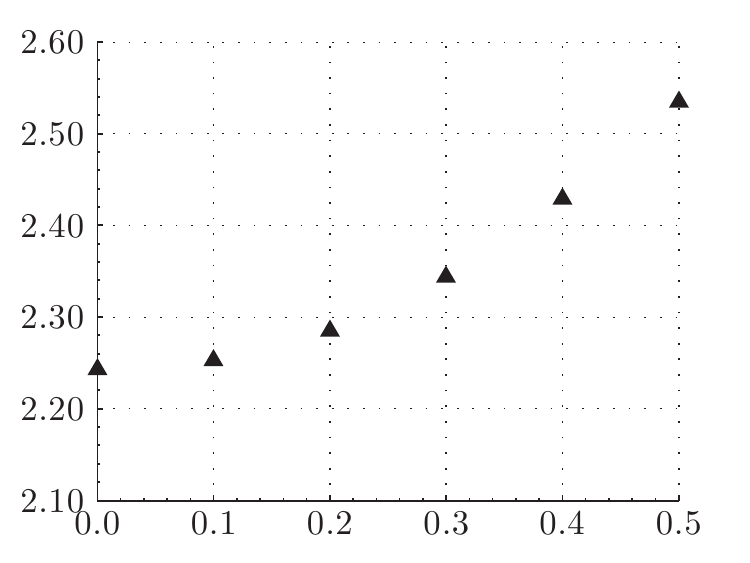} % see overpic for details.
    \end{overpic}} \\
    & \center{$M_{\infty}$} & & \center{$M_{\infty}$} & \\
  \end{tabular}
  \caption{Wake characteristics for $0 \le M_\infty \le 0.5$ and $Re = 20$ and $40$ (steady flow).}
  \label{fig:wake_comp}
  \end{center}
\end{figure}

We first examine the effect of compressibility on the characteristic closed wake dimensions for stable flows. Shown in figure \ref{fig:steady_vbarmag} are contours of normalized velocity magnitude $\|\textbf{u}\|/u_{\infty}$ in the near field of the cylinder. These flow fields indicate that the wake elongates as $M_{\infty}$ is increased from $0$ to $0.5$. A similar trend has been established in the literature for incompressible flows with increasing $Re$ \citep{Coutanceau:JFM77a}. To further investigate the influence of compressibility on the wake, normalized wake geometry parameters are presented as functions of $M_{\infty}$ for various stable $Re$ in figure \ref{fig:wake_comp}.  From these measurements, it can be seen that the wake increases in length (as measured by $l_r/d$).  It is also notable that over the range of $M_{\infty}$ examined, the variation in $l_a/d$ and $l_r/d$ relative to their incompressible values is more than twice as large at $Re = 40$ compared to the variation observed at $Re = 20$. This result indicates that as $Re$ becomes further removed from $Re_c$, the wake becomes less susceptible to compressibility effects for stable flows. A similar result is obtained in terms of $St$ for unstable flows and is discussed in further detail later in this section.

\begin{figure}
  \begin{center}
  \begin{tabular}{m{0.009\textwidth}m{0.48\textwidth}m{0.009\textwidth}m{0.48\textwidth}m{0.0001\textwidth}}
    \rotatebox[origin=c]{90}{$\overline{C}_d$} &
    {\scriptsize
    \begin{overpic}[width=0.48\textwidth]{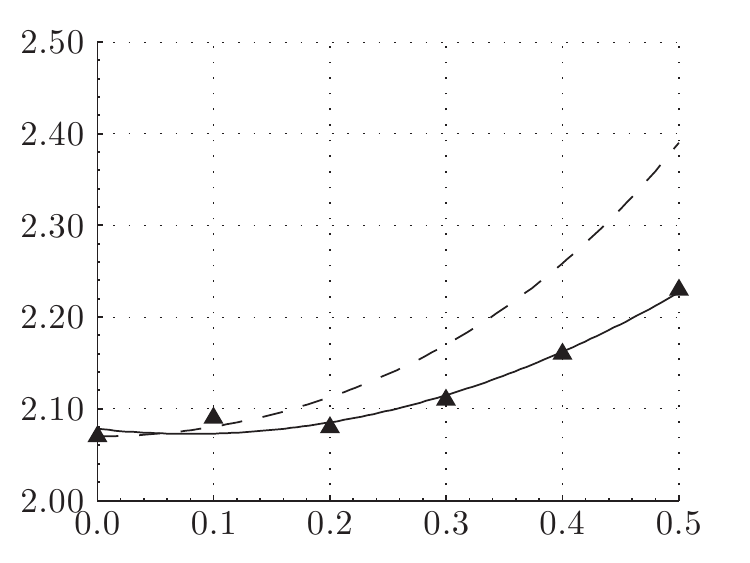} % see overpic for details.
      \put(20,60){$Re = 20$}
    \end{overpic}} & 
    \rotatebox[origin=c]{90}{$\overline{C}_d$} &
    {\scriptsize
    \begin{overpic}[width=0.48\textwidth]{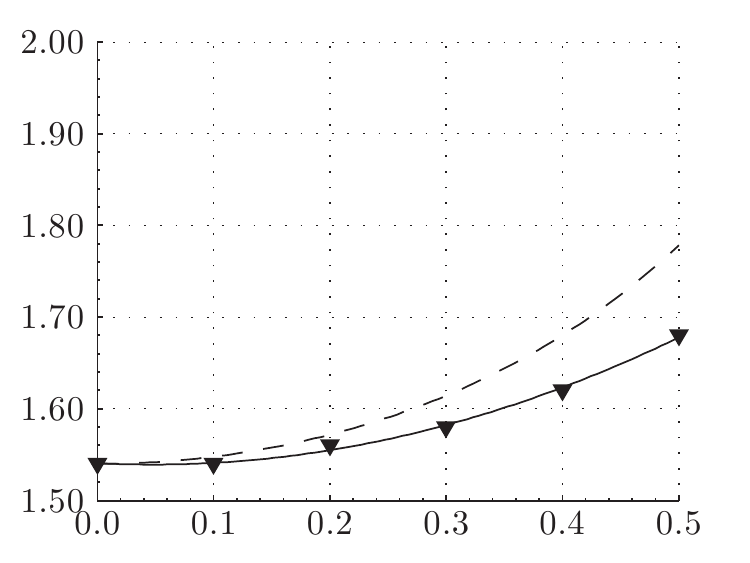} % see overpic for details.
      \put(20,60){$Re = 40$}
    \end{overpic}} & \\
    \rotatebox[origin=c]{90}{$\overline{C}_d$} &
    {\scriptsize
    \begin{overpic}[width=0.48\textwidth]{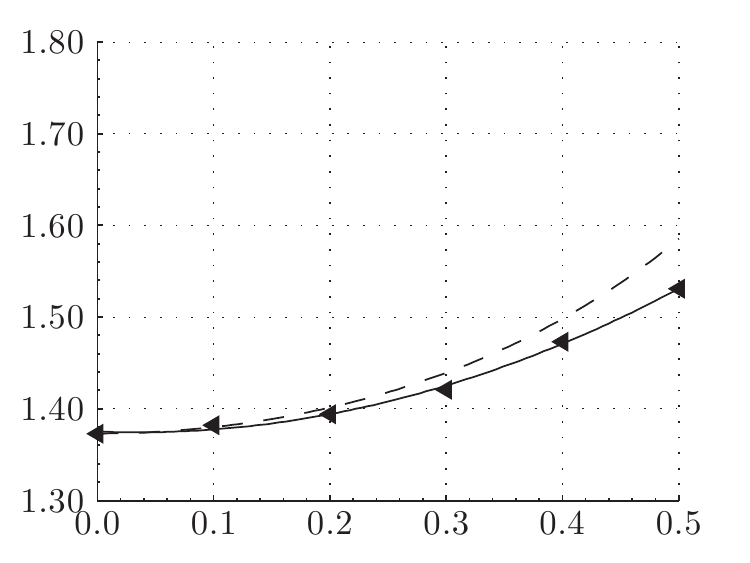} % see overpic for details.
      \put(20,60){$Re = 75$}
    \end{overpic}} & 
    \rotatebox[origin=c]{90}{$\overline{C}_d$} &
    {\scriptsize
    \begin{overpic}[width=0.48\textwidth]{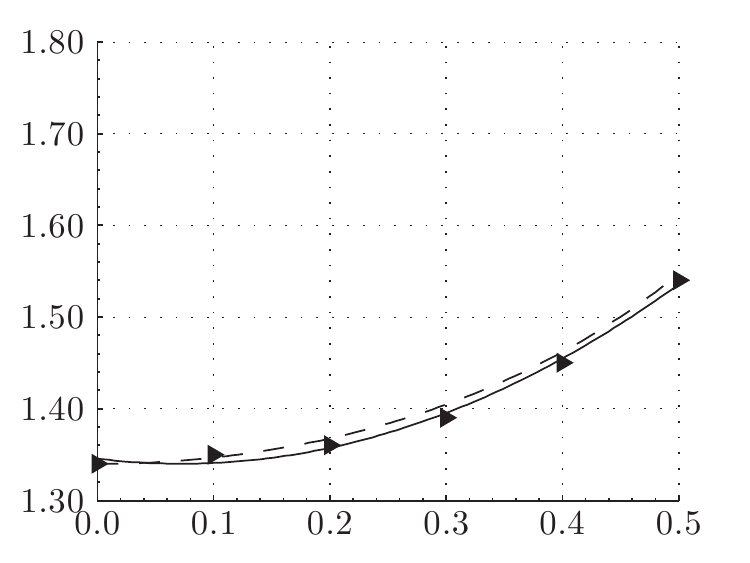} % see overpic for details.
      \put(20,60){$Re = 100$}
    \end{overpic}} & \\
    & \center{$M_{\infty}$} &
    & \center{$M_{\infty}$} & \\
  \end{tabular}
  \caption{Drag coefficient $\overline{C}_d$ for selected stable and unstable cases. Overlaid solid lines are quadratic curve fits of the present data, while the dashed lines are the prediction given by the Prandtl--Glauert transformation.}
  \label{fig:Cd_norm}
  \end{center}
\end{figure}

\begin{figure}
  \begin{center}
  \begin{tabular}{m{0.009\textwidth}m{0.48\textwidth}m{0.009\textwidth}m{0.48\textwidth}m{0.0001\textwidth}}
    & \center{$Re = 20$} & 	& \center{$Re = 40$} & \\
    \rotatebox[origin=c]{90}{$C_p$} & 
    {\scriptsize
    \begin{overpic}[width=0.48\textwidth]{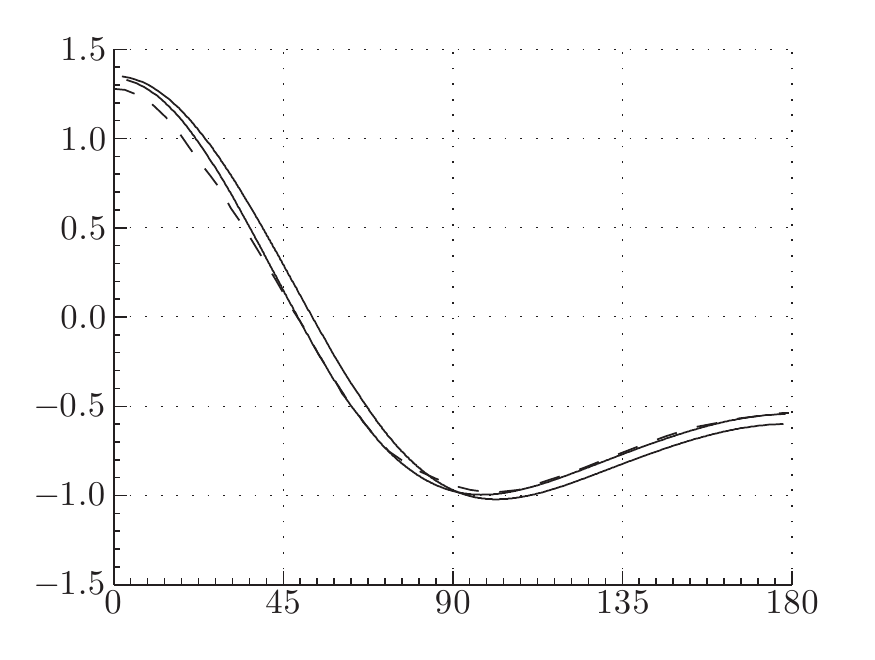} % see overpic for details.
       \put(18,28){$M_\infty = 0.0$} % add grid to [grid, width=...] above to see coordinates
       \put(38,40){$M_\infty = 0.5$} 
    \end{overpic}} &	
    \rotatebox[origin=c]{90}{$C_p$} & 
    {\scriptsize
    \begin{overpic}[width=0.48\textwidth]{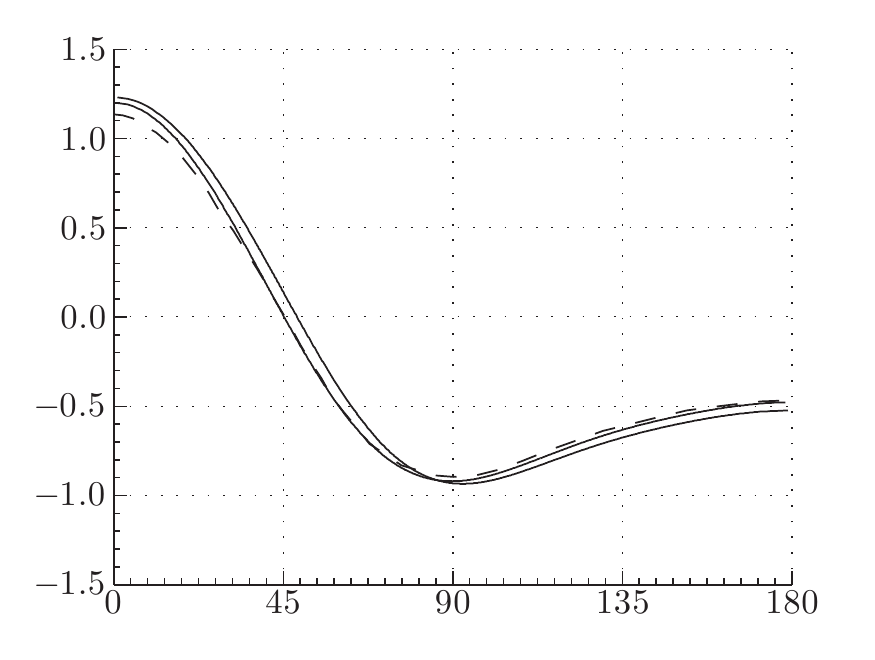} % see overpic for details.
       \put(16,28){$M_\infty = 0.0$} % add grid to [grid, width=...] above to see coordinates
       \put(36,40){$M_\infty = 0.5$} 
    \end{overpic}} & \\
    & \center{$\theta$ (deg)} &	& \center{$\theta$ (deg)} &  \\
  \end{tabular}
  \caption{$C_p$ distributions over the cylinder surface. Solid lines are from the present study. Overlaid in dashed lines are the incompressible solutions from \cite{Fornberg:JFM80}.}
  \label{fig:cpavg}
  \end{center}
\end{figure}

Based on observations from the steady wakes given in figures \ref{fig:steady_vbarmag} and \ref{fig:wake_comp}, it is expected that $\overline{C}_d$ should increase along with $M_{\infty}$, as the wake exhibits a clear increase in size. This expectation is further justified by the previous observation of such behaviour for airfoils in similar flow conditions \citep{Munday:JA15}. Indeed, as shown in figure \ref{fig:Cd_norm}, we observe an increase in $\overline{C}_d$ coincident with an increase in $M_{\infty}$. To supplement the explanation for this variation in $\overline{C}_d$ given by the wakes, we present distributions of $C_p$ along the cylinder surface for selected cases in figure \ref{fig:cpavg}. From these distributions, we find that as $M_{\infty}$ increases, the pressure along the front half of the cylinder (i.e., $\theta < 90^\circ$) increases while the pressure at the rear stagnation point decreases. When coupled with the aforementioned shift in wake geometry, this increase in the pressure gradient between the upstream and downstream halves of the cylinder explains the observed rise in $\overline{C}_d$ with $M_{\infty}$ for stable cases.

\begin{figure}
  \begin{center}  
  \begin{tabular}{m{0.13\textwidth}m{0.43\textwidth}m{0.43\textwidth}m{0.0001\textwidth}} \hline 
   &
   \center{$\omega d / u_{\infty}$} &
   \center{$\|\overline{\textbf{u}}\|/u_{\infty}$} & \\ \hline
   $M_{\infty} = 0$ & 
   \includegraphics[trim = 0cm 0cm 9cm 0cm, clip = true, width=0.4\textwidth]{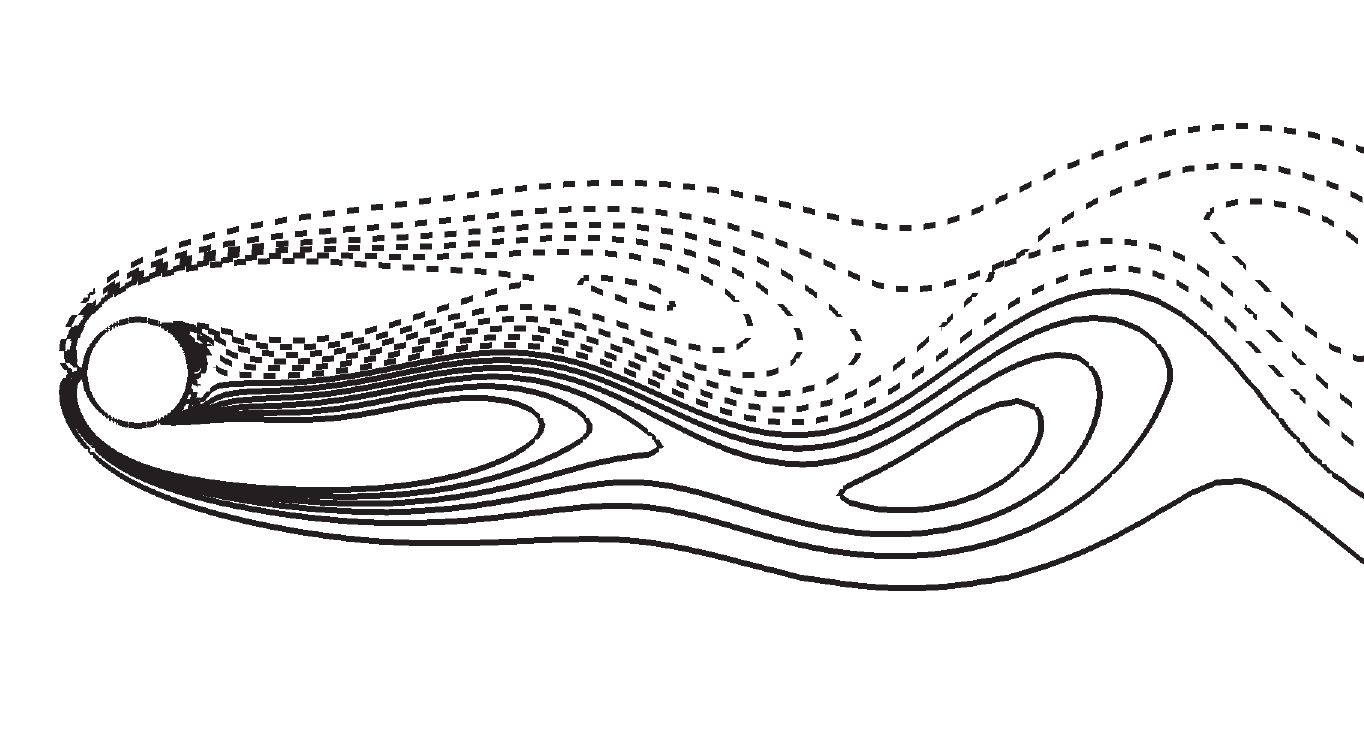} & 
   \includegraphics[trim = 0cm 0cm 9cm 0cm, clip = true, width=0.4\textwidth]{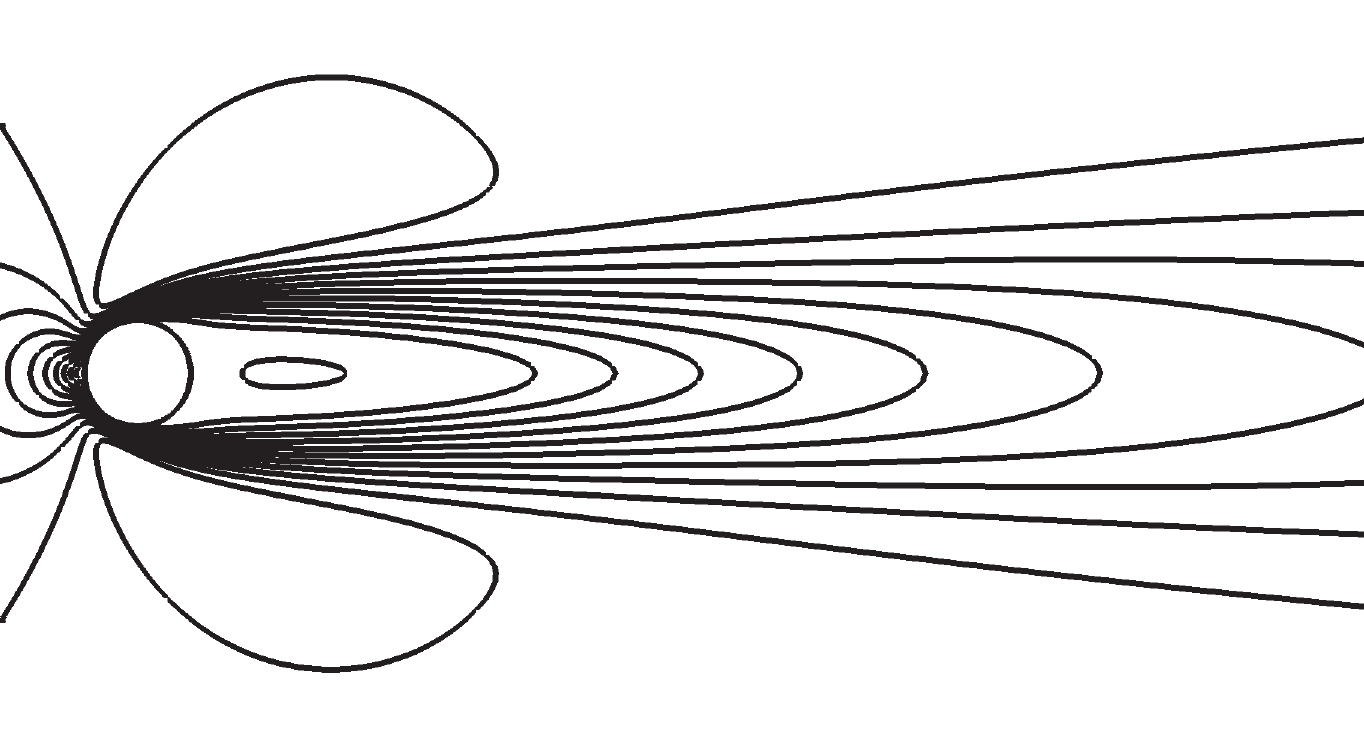} \\
   $M_{\infty} = 0.3$ &
   \includegraphics[trim = 0cm 0cm 9cm 0cm, clip = true, width=0.4\textwidth]{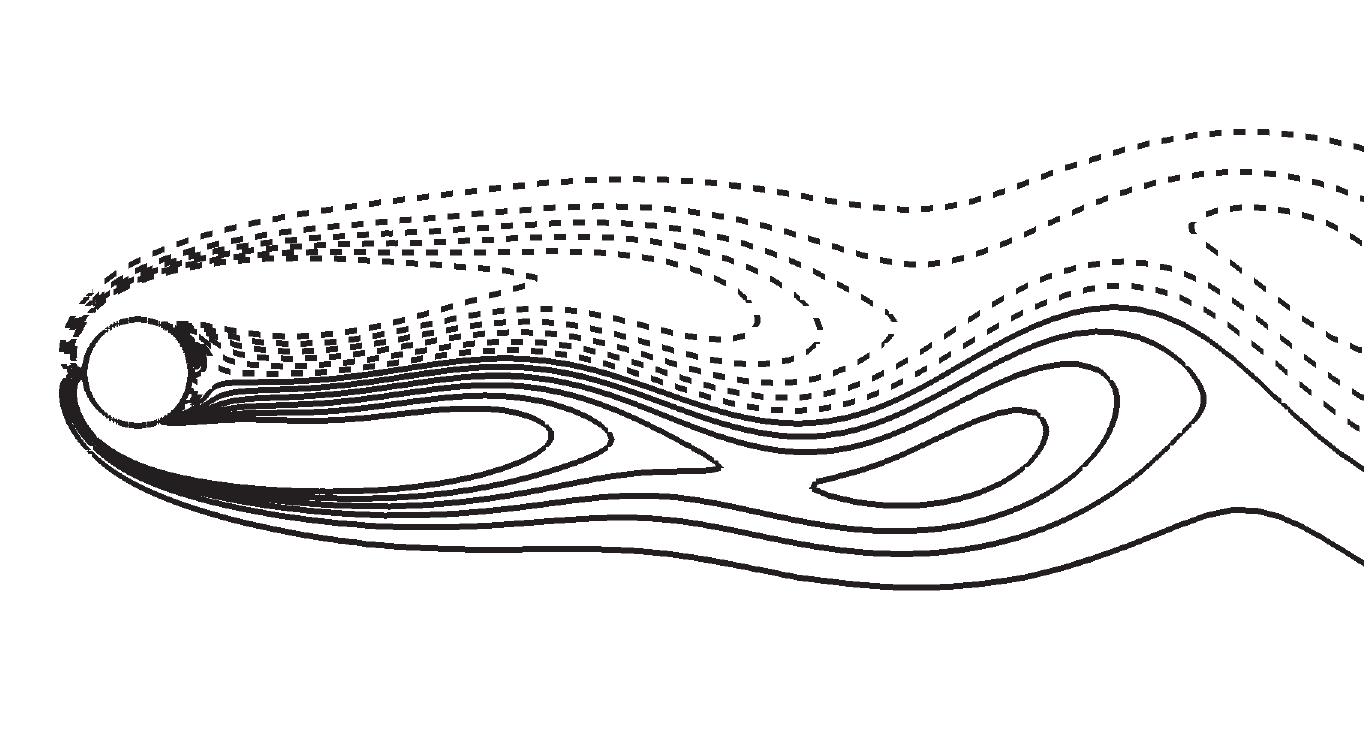} &
   \includegraphics[trim = 0cm 0cm 9cm 0cm, clip = true, width=0.4\textwidth]{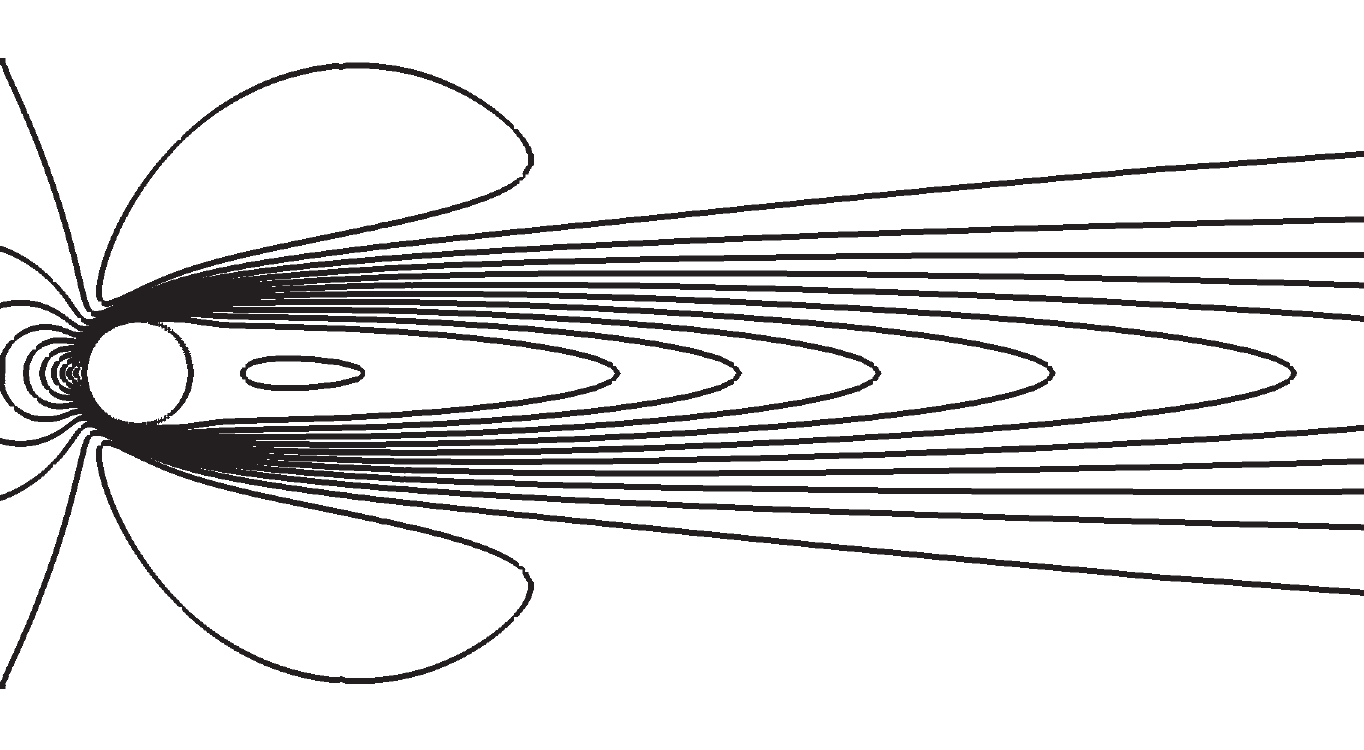} \\
   $M_{\infty} = 0.5$ &
   \includegraphics[trim = 0cm 0cm 9cm 0cm, clip = true, width=0.4\textwidth]{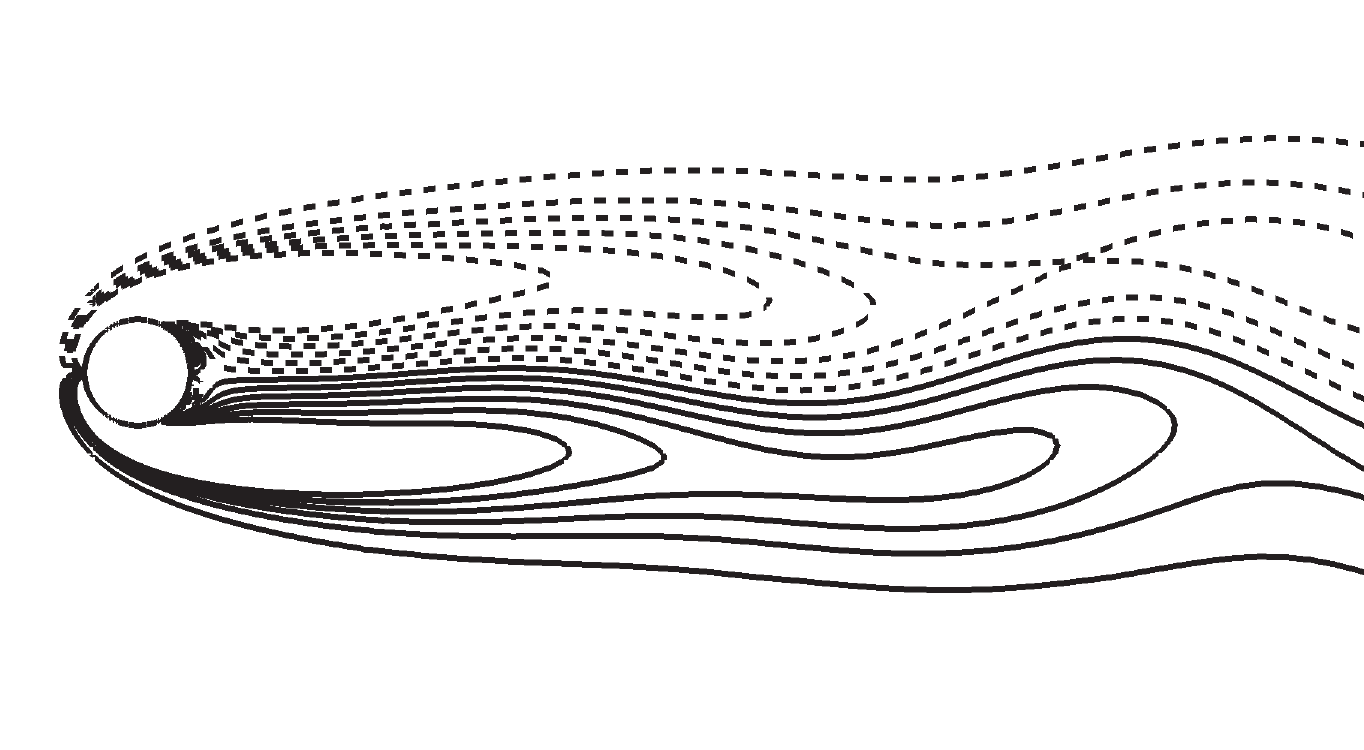} &
   \includegraphics[trim = 0cm 0cm 9cm 0cm, clip = true, width=0.4\textwidth]{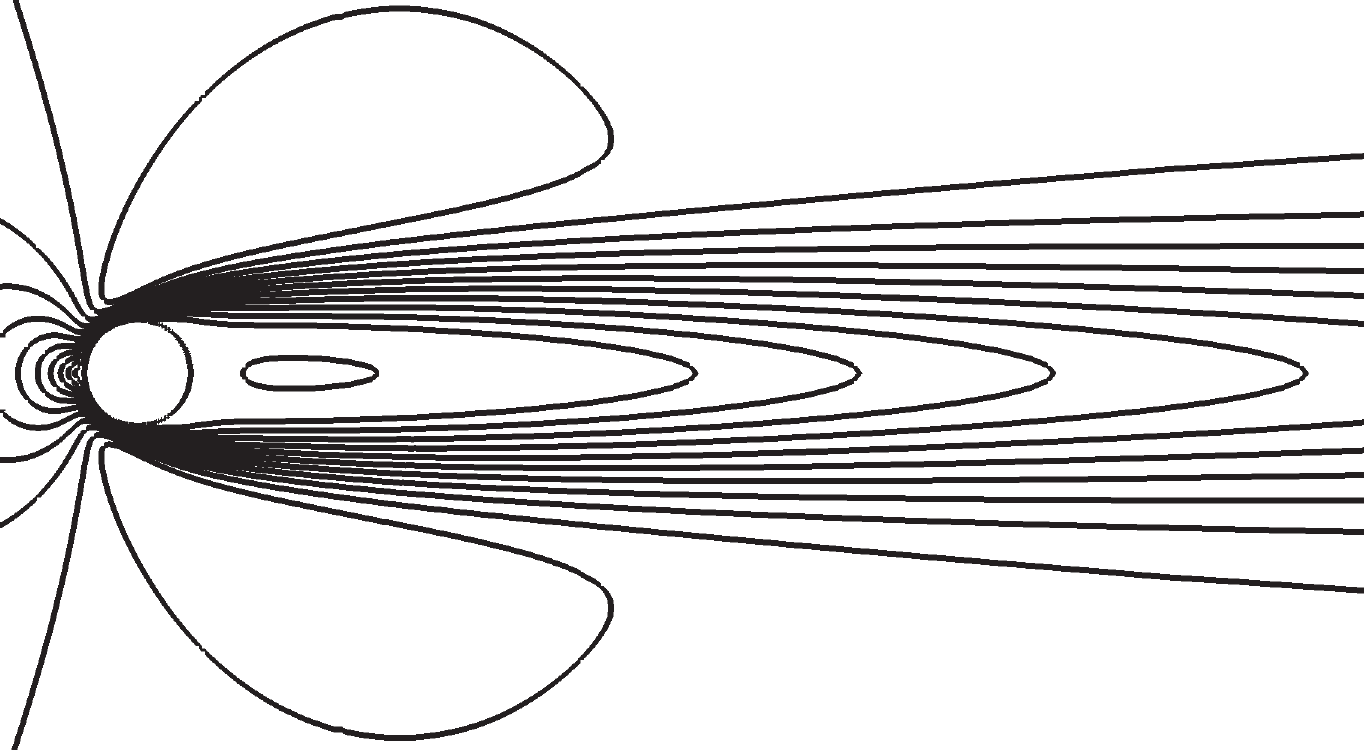} \\ \hline %\trim goes CCW from left side
  \end{tabular} 
  \caption{Flow field comparisons for varying $M_{\infty}$ at $Re = 50$. Vorticity contours have 12 levels between $\pm M_{\infty}$, with dashed lines indicating negative values. Time-averaged velocity contours have 12 levels between 0 and 1.1.}
  \label{fig:R50_comparison}
  \end{center}
\end{figure}

Interestingly, in comparing $\overline{C}_d$ for stable and unstable cases, we find that both classes of flow exhibit quadratic variations with $M_{\infty}$, as indicated by the solid lines in figure \ref{fig:Cd_norm}. To explain this behaviour for unstable cases, we examine the flowfields obtained at $Re = 50$ for varying $M_{\infty}$ in figure \ref{fig:R50_comparison}. From the contours of time-averaged velocity magnitude $\|\overline{\textbf{u}}\|/u_{\infty}$, it can be seen that the effect of increasing $M_{\infty}$ is to increase the size of the time-averaged recirculation region. This effect matches the increase in wake size with $M_{\infty}$ previously displayed in figure 3 for stable flows, and thus explains the parallel trends in $\overline{C}_d$.

It is worth mentioning that the trends in $\overline{C}_d$ behave somewhat similarly to the well-known Prandtl--Glauert transformation \citep{Glauert:RSPA28}.  While this transform holds only for lift force in inviscid flow, we observe that for higher Reynolds number flows in this study, the Prandtl--Glauert type transformation is also able to approximate the drag force for compressible flow based on incompressible drag values 
\begin{equation}
   C_d = \frac{C_{d,M_{\infty} = 0}}{\sqrt{1 - M_{\infty}^2}} = C_{d,M_{\infty} = 0} \left( 1 + \frac{1}{2}M_\infty^2 + \frac{3}{8} M_\infty^4 + \cdots \right),
  \label{eqn:PG}
\end{equation}
as shown in figure {\ref{fig:Cd_norm}} by the dashed lines. It is expected that the Prandtl--Glauert transformation does not accurately predict the compressibility effects on a cylinder since the transformation is also based on small deflections of the freestream as it flows over a slender body. However, the data appear to approach the theoretical trend with increasing $Re$ (e.g., $Re = 100$). We attribute this result to the fact that the Prandtl--Glauert type transformation performs well for flows at higher $Re$, for which viscous effects are of small magnitude compared to pressure effects.

\begin{figure}
  \begin{center}
  \begin{tabular}{m{0.009\textwidth}m{0.48\textwidth}m{0.0001\textwidth}}
    \rotatebox[origin=c]{90}{$St$}	& 
    {\scriptsize
    \begin{overpic}[width=0.48\textwidth]{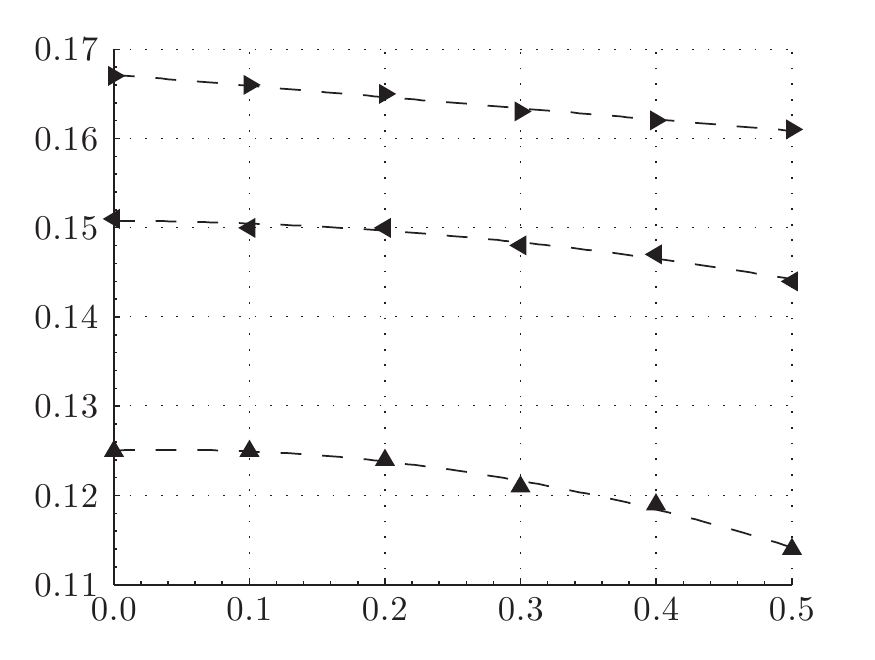}
      \put(71,64){$Re = 100$}
      \put(68,50){$Re = 75$}
      \put(63,21){$Re = 50$}
    \end{overpic}} &  \\
    & \center{$M_{\infty}$} & \\
  \end{tabular}
  \caption{$St$ as a function of $M_{\infty}$ for various unstable $Re$.}
  \label{fig:reducedSt_M}
  \end{center}
\end{figure}

Though our discussion to this point has focused on the effects of compressibility on steady or time-averaged observations, changes in $M_{\infty}$ are also reflected in unsteady flow phenomena. For instance, the snapshots of non-dimensionalized spanwise vorticity $\omega d / u_{\infty}$ in figure 7 indicate that compressibility tends to increase the wavelength of vortex shedding, thus leading to the aforementioned elongation of the time-averaged recirculation region.  Besides shedding wavelength, the effect of $M_{\infty}$ on shedding frequency can be seen in figure \ref{fig:reducedSt_M}. Shown in this figure is the Strouhal number $St$ as a function of $M_{\infty}$ for different $Re$. From this plot, it is evident that increasing $M_{\infty}$ reduces the frequency of vortex shedding. This trend matches the increase in vortex shedding wavelength previously mentioned. Moreover, we see that the data shows a smaller change in $St$ from $M_\infty = 0$ to $0.5$ as $Re$ is increased beyond $Re_c$, with over a $9\%$ reduction at $Re = 50$, but only about $4\%$ at both $Re = 75$ and $100$. This result indicates that $St$ exhibits reducing sensitivity to compressibility effects as $(Re - Re_c)$ increases. However, we have only made this observation for $47 \le Re \le 100$. Further investigation is needed to determine whether it holds up to and beyond the limit for spanwise instability ($Re \approx 190$) \citep{Williamson:ARFM96}.  Nonetheless, it is notable that this behavior parallels that already discussed for the characteristic wake parameters in stable flow.

\begin{figure}
  \begin{center}  
  \begin{tabular}{m{0.10\textwidth}m{0.43\textwidth}m{0.43\textwidth}m{0.0001\textwidth}} \hline
			& \center{$\omega d / u_{\infty}$}																	& \center{$\|\overline{\textbf{u}}\|/u_{\infty}$}	&	\\ \hline
  $Re = 60$ 	& \includegraphics[trim = 0cm 0cm 9cm 0cm, clip = true, width=0.4\textwidth]{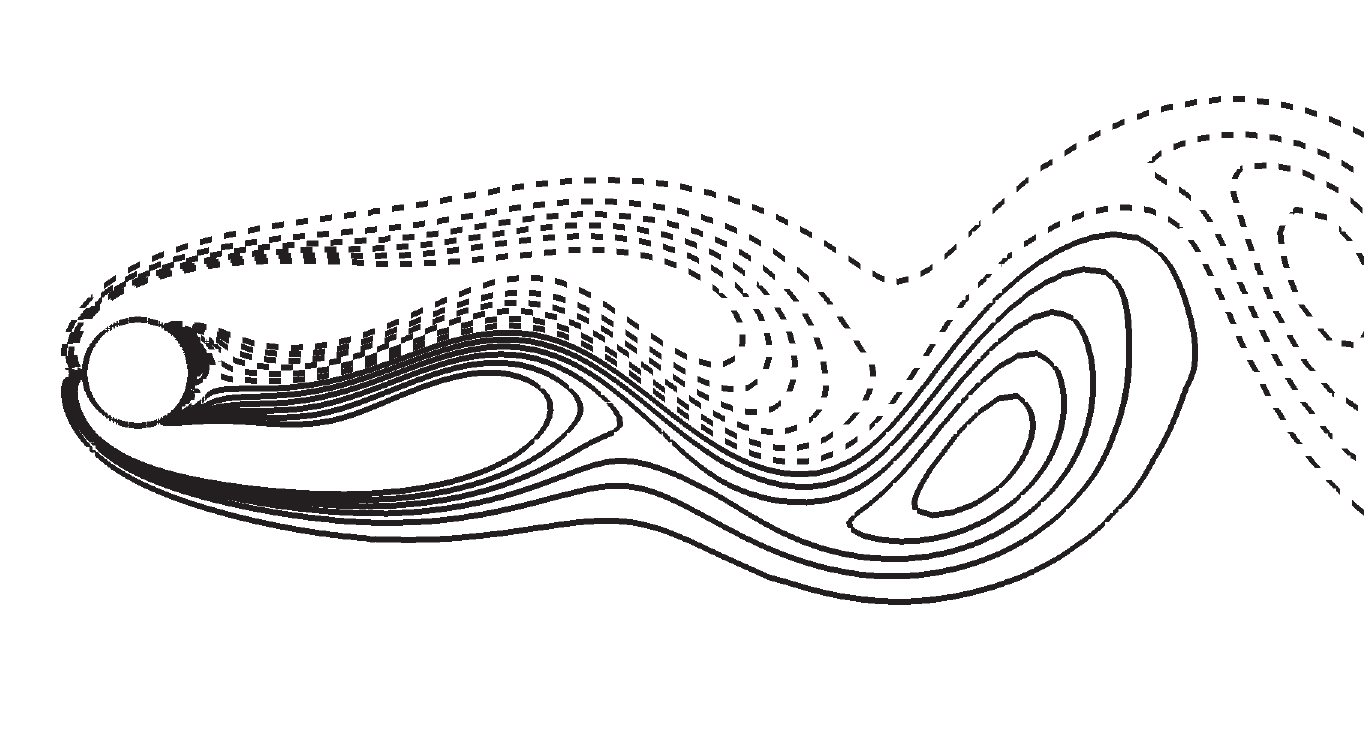} 		& \includegraphics[trim = 0cm 0cm 9cm 0cm, clip = true, width=0.4\textwidth]{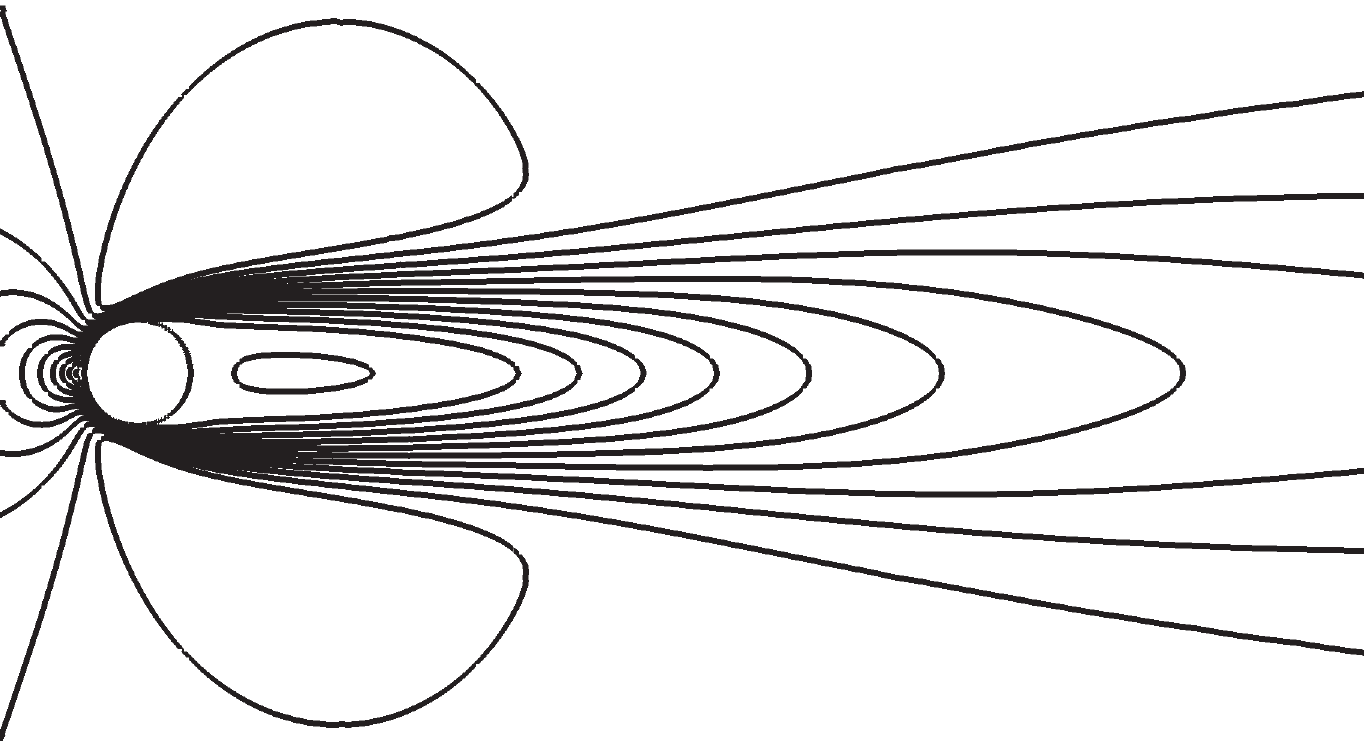} 		\\
  $Re = 80$ 	& \includegraphics[trim = 0cm 0cm 9cm 0cm, clip = true, width=0.4\textwidth]{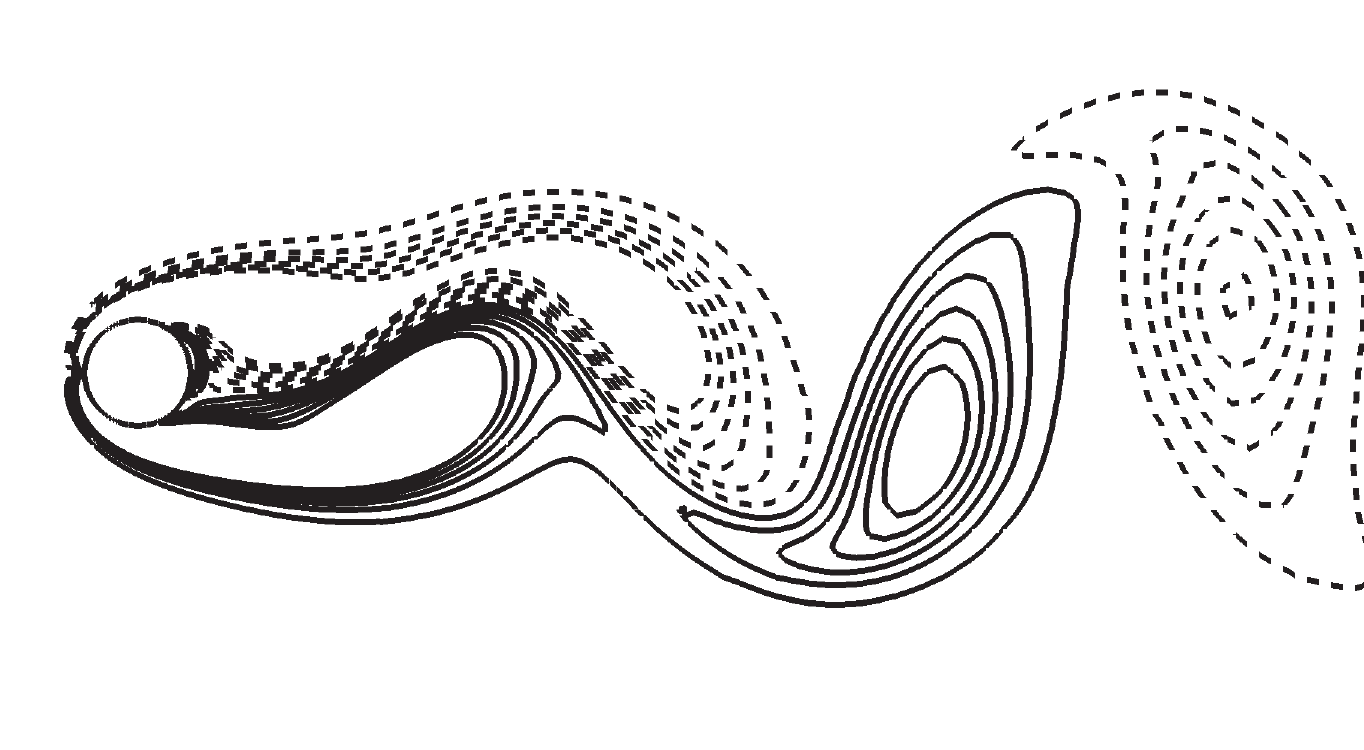} 		& \includegraphics[trim = 0cm 0cm 9cm 0cm, clip = true, width=0.4\textwidth]{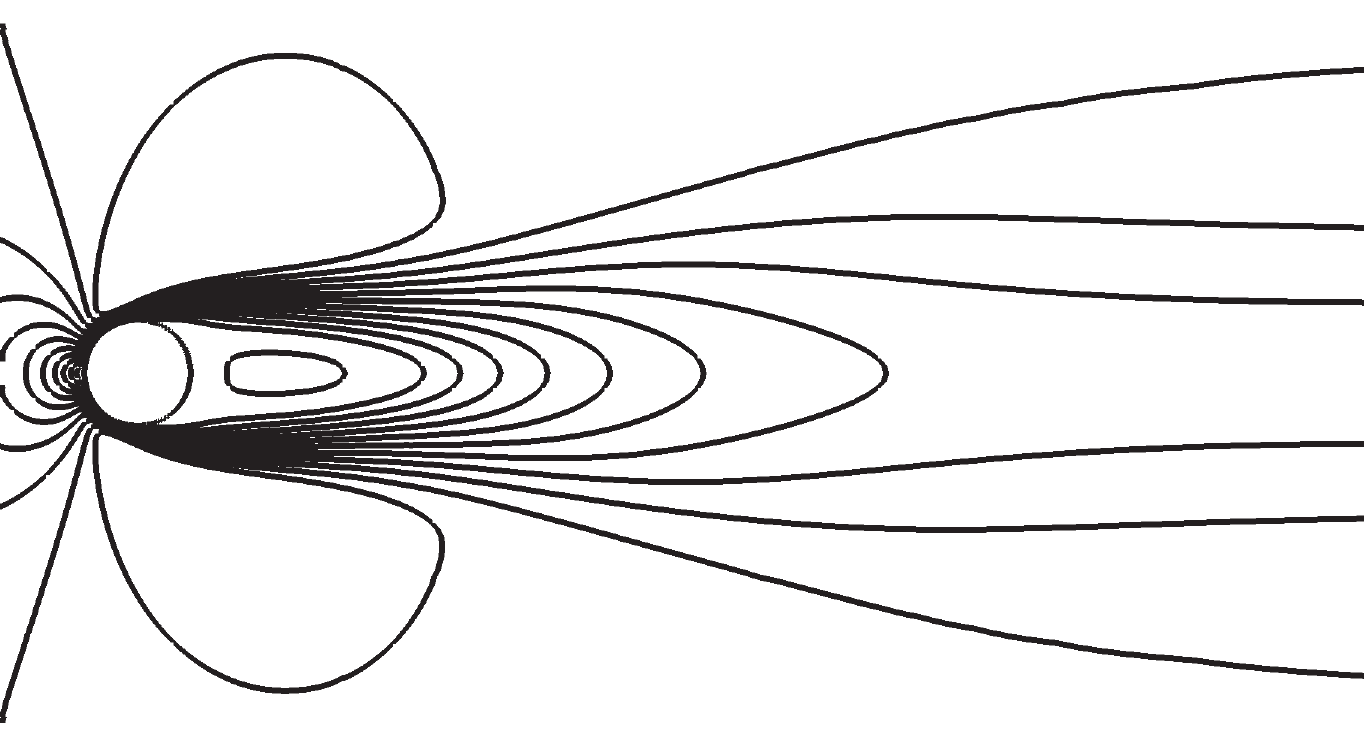} 		\\
  $Re = 100$ 	& \includegraphics[trim = 0cm 0cm 9cm 0cm, clip = true, width=0.4\textwidth]{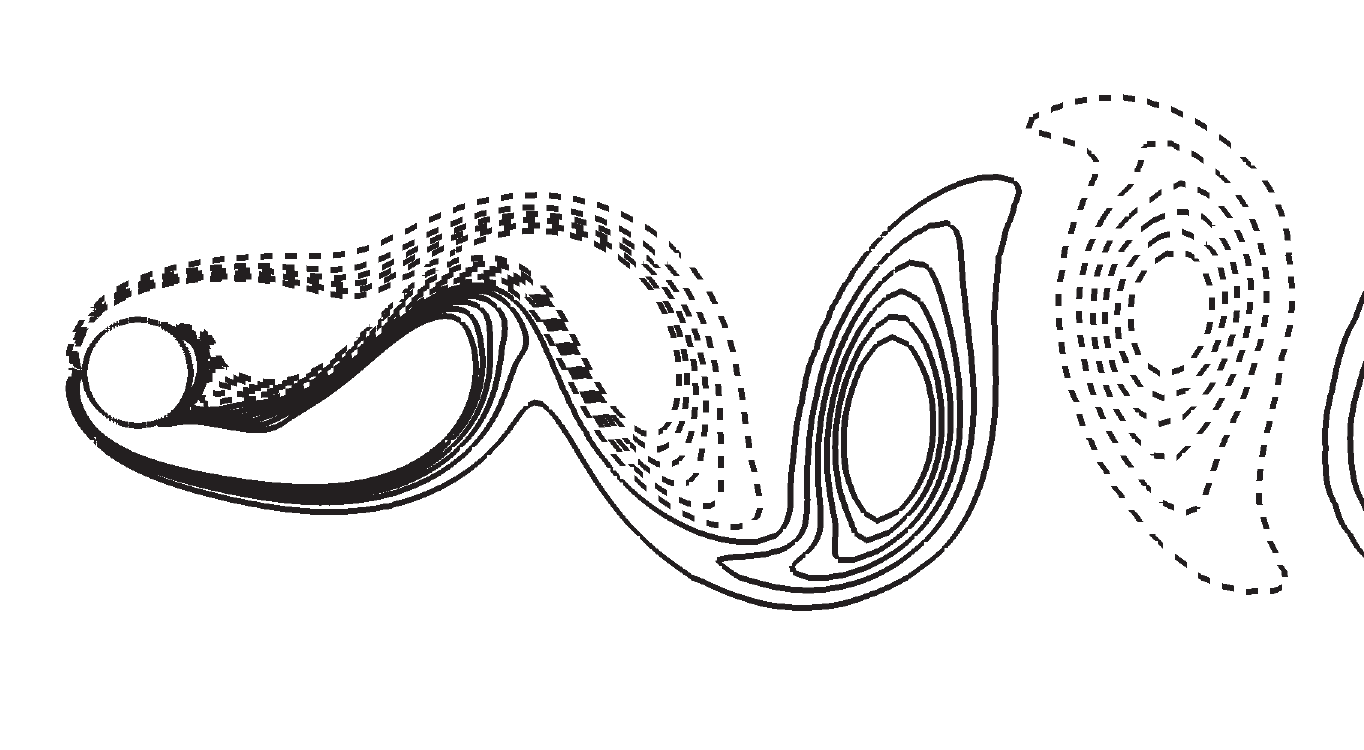} 	& \includegraphics[trim = 0cm 0cm 9cm 0cm, clip = true, width=0.4\textwidth]{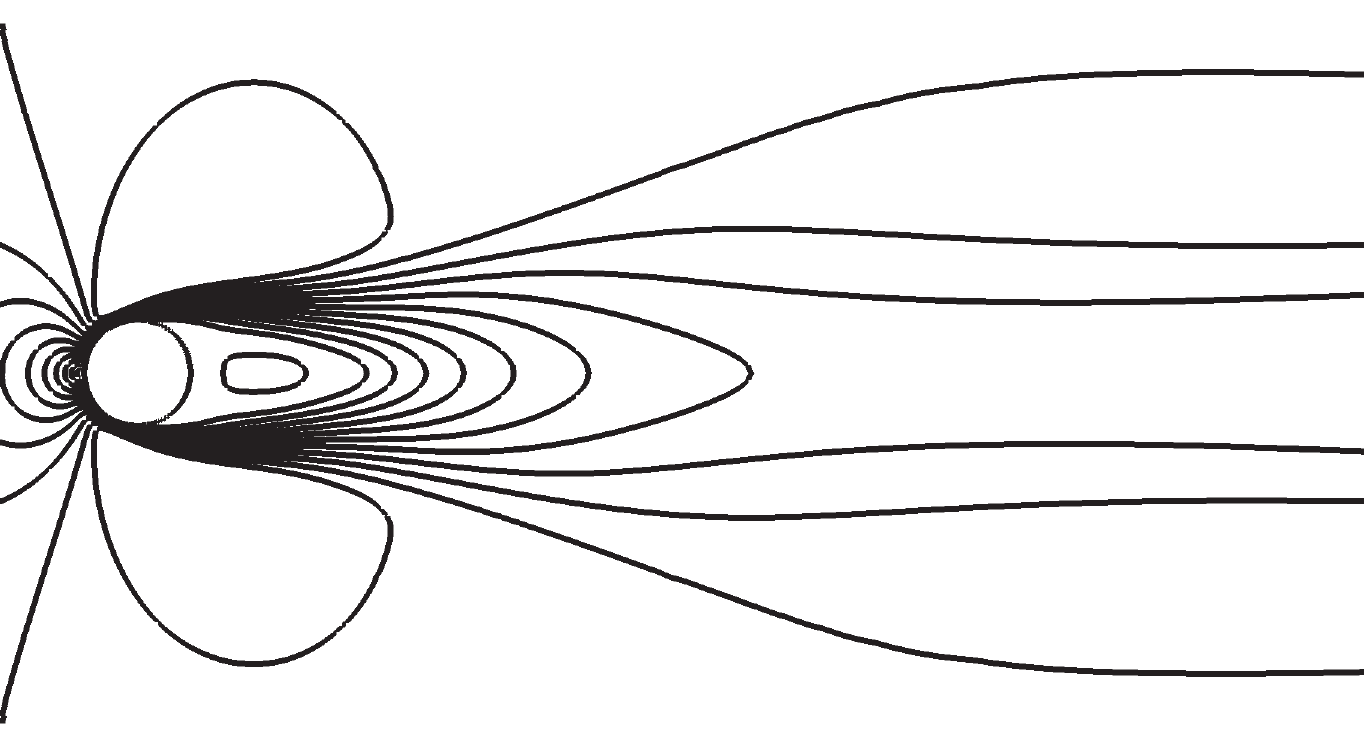} 	\\ \hline 
  \end{tabular} 
  \caption{Flow field comparisons for varying $Re$ at $M_{\infty} = 0.5$. Vorticity contours have 12 levels between $\pm M_{\infty}$, with dashed lines indicating negative values. Time-averaged velocity contours have 12 levels between 0 and 1.1.} 
  \label{fig:M05_comparison}
  \end{center}
\end{figure} 

Finally, to isolate the effects of varying $Re$ above $Re_c$ for flows above the incompressible limit ($M_{\infty} \approx 0.3$), figure \ref{fig:M05_comparison} shows the non-dimensionalized spanwise vorticity $\omega d / u_{\infty}$ and velocity magnitude $\|\overline{\textbf{u}}\|/u_{\infty}$ for different $Re$ at $M_{\infty} = 0.5$. From the flow fields, it can be seen that with increasing $Re$, both the vortex-shedding length and the time-averaged recirculation region shorten. This trend has already been well-established for unsteady incompressible flows near $Re_c$ \citep{Coutanceau:JFM77a}, and is now shown here to hold above the incompressible limit as well.

%%%%%%%%%%%%%%%%%%%%%%%%%%%%%%%%%%%%%%%%%%%%%%%%%%%%%%%%%%%%%%%%%%

\subsection{Linear stability analysis}

The generation of wake vortices and the vortex street that appears in unstable flow over a circular cylinder gives rise to unsteady forces. To better understand the formation of the vortex street in compressible flow, we examine the growth of linear instability for unstable flows. 

\begin{figure}
  \begin{center}
  \begin{tabular}{m{0.13\textwidth}m{0.43\textwidth}m{0.43\textwidth}m{0.0001\textwidth}} \hline
  & \center{time-averaged flow} & \center{unstable steady state} & \\ \hline
  $\|\overline{\textbf{u}}\|/u_{\infty}$ &
  \includegraphics[trim = 0cm 0cm 9cm 0cm, clip = true, width=0.4\textwidth]{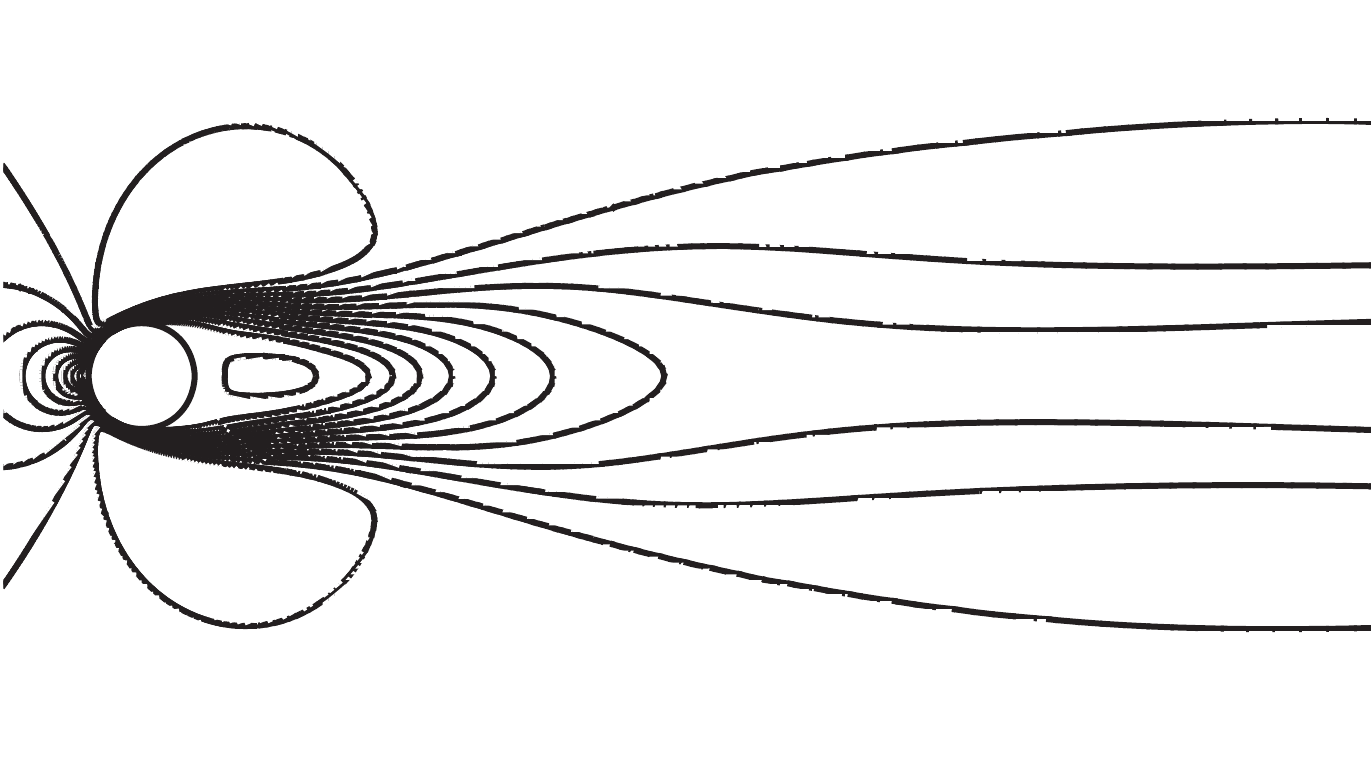} & 
  \includegraphics[trim = 0cm 0cm 9cm 0cm, clip = true, width=0.4\textwidth]{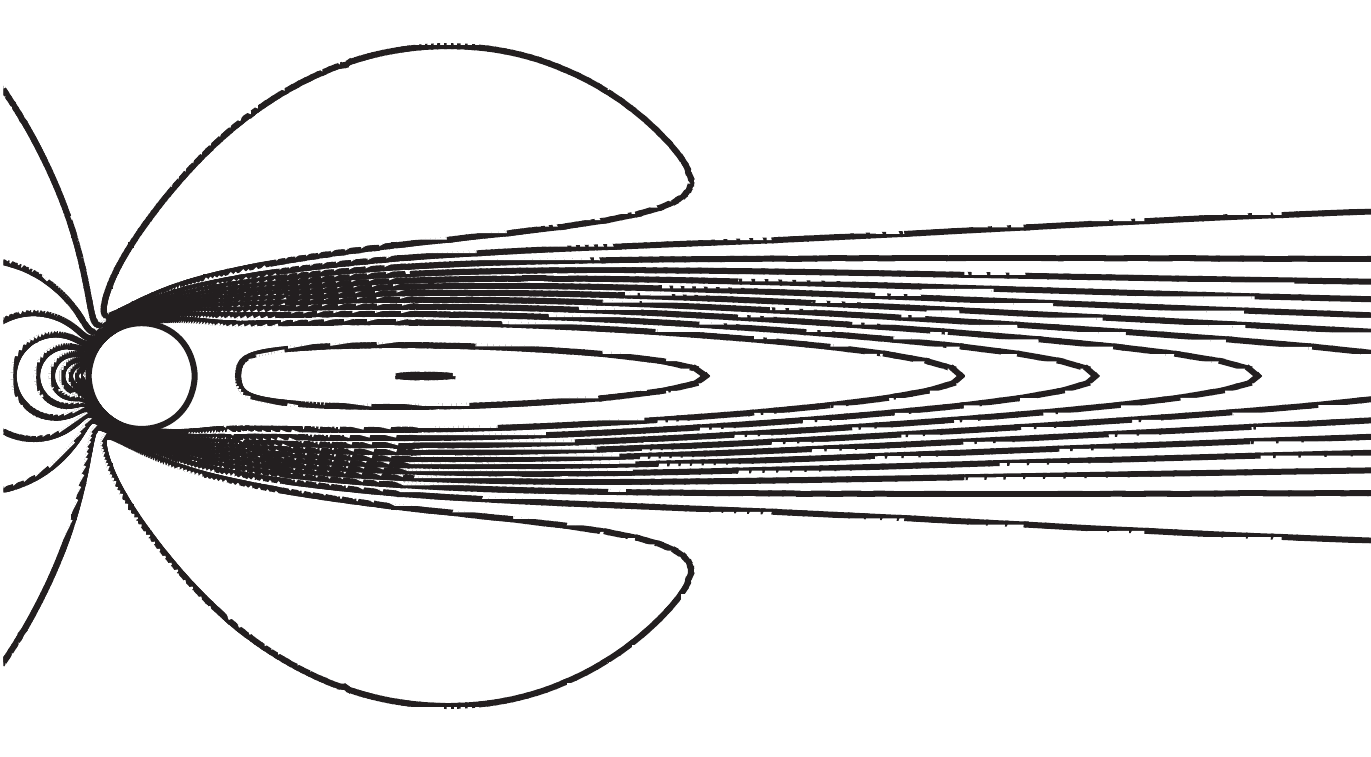} \\
  $\omega d / u_{\infty}$ & 
  \includegraphics[trim = 0cm 0cm 9cm 0cm, clip = true, width=0.4\textwidth]{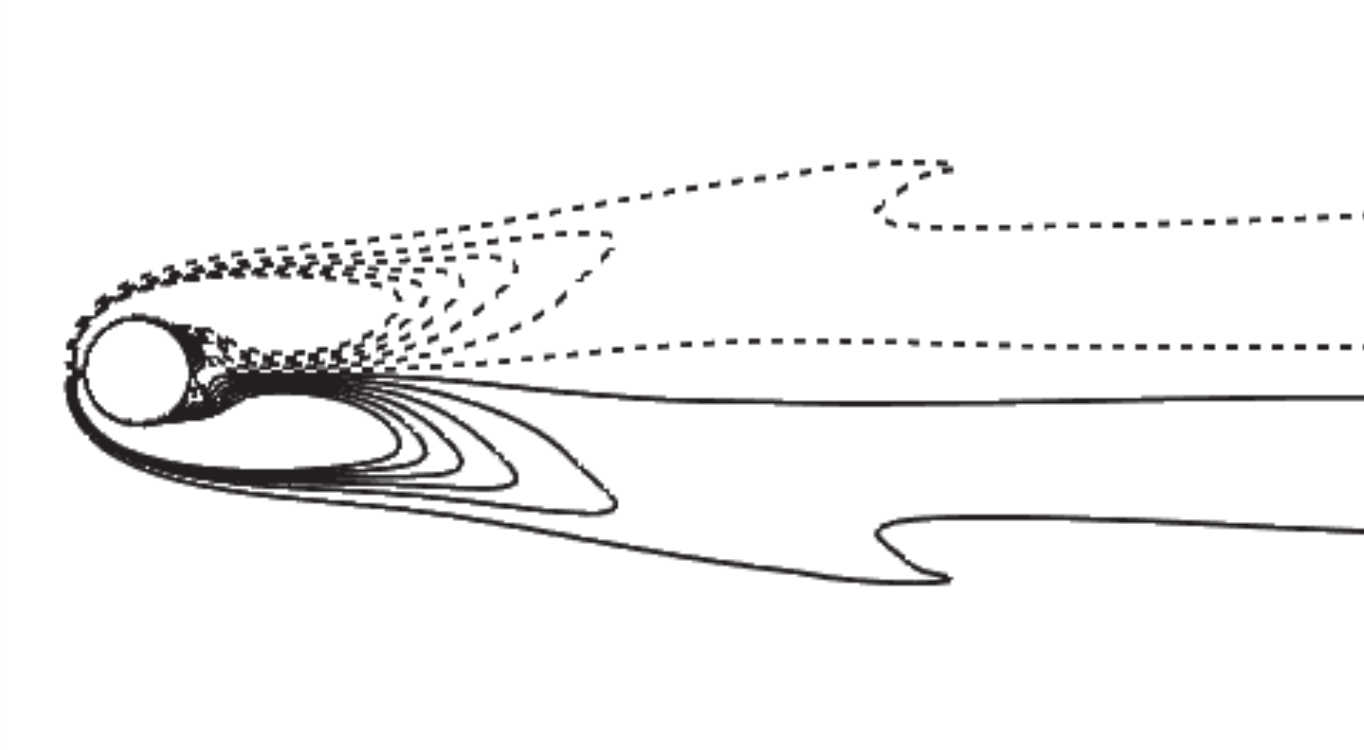} & 
  \includegraphics[trim = 0cm 0cm 9cm 0cm, clip = true, width=0.4\textwidth]{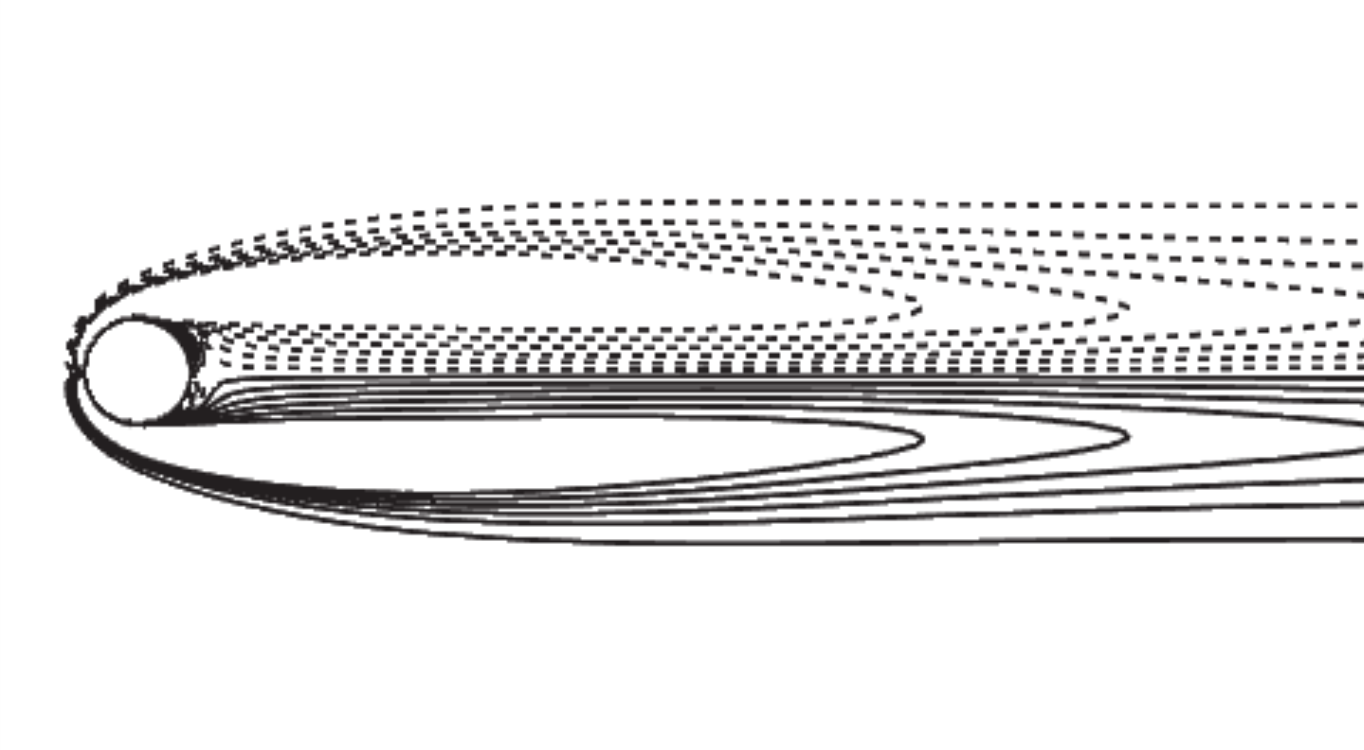} \\ \hline
  \end{tabular} 
  \caption{Comparison of velocity and vorticity fields for the time-averaged flow and unstable steady states for $Re = 100$ and $M_\infty = 0$.  Contour settings follow Figs.~\ref{fig:M05_comparison} and \ref{fig:R50_comparison}.}
  \label{fig:USS}
  \end{center}
\end{figure}

To perform linear stability analysis, we first find the base state from which to perturb the flow.  The unstable steady state is used as the base state and is numerically determined for each combination of $Re>Re_c$ and $M_\infty$ using the selective frequency damping method \citep{Akervik:PF06}.  This solver technique essentially damps out any oscillation in the flow field using proportional feedback control with temporal filtering to force the flow to the unstable steady state.  We note that this unstable steady state is not the time-averaged state, as shown for an example with $Re = 100$ and $M_\infty = 0$ in figure~\ref{fig:USS}.  The most immediately recognizable difference between the two flows is the sharp kink seen in the vorticity field of the time-averaged flow, which is absent from the smooth profile of the unstable steady state.  Also, note that the unstable steady state for flow over a circular cylinder matches the steady symmetric profile examined by \cite{Fornberg:JFM80} for incompressible flow.

The present linear stability analysis is conducted by performing numerical simulation using the unstable steady state taken as the initial condition with small numerical perturbation introduced.  We emphasize that while the simulation is performed with the full nonlinear Navier--Stokes equations, the nonlinear advection term is negligible for the considered level of perturbation. Furthermore, by limiting our focus to the time period just after the introduction of the perturbation, the magnitude of $C_l$ is assumed to be sufficiently small to allow us to ignore the nonlinear term in (\ref{eqn:SL}). We thus perform linear stability analysis by tracking the time evolution of the perturbation to the lift coefficient $C_l$ through the linearized version of (\ref{eqn:SL})
\begin{equation}
  \frac{\mathrm{d}C_l}{\mathrm{d}t} = \sigma C_l
  \label{eqn:SLCL}
\end{equation}
for the dominant instability. The solution to this equation is
\begin{equation}
  C_l(t) = C_{l,t=0}\textrm{e}^{\sigma t},
  \label{eqn:CLT}
\end{equation}
where $C_{l,t=0}$ is the initial lift amplitude generated by a small perturbation (note that lift for the unstable steady state is zero). By fitting curves of the form given in (\ref{eqn:CLT}) to $C_l$ values computed during simulation, the growth rate $\sigma_r =\textrm{Re}(\sigma)$ and dominant unstable frequency $\sigma_i =\textrm{Im}(\sigma)$ are determined for the chosen $Re$ and $M_{\infty}$. This procedure without the determination of unstable steady state has been used previously by \cite{Bres:JFM08} for tracking the decay of disturbances for linear stability analysis of compressible flow over open rectangular cavities.

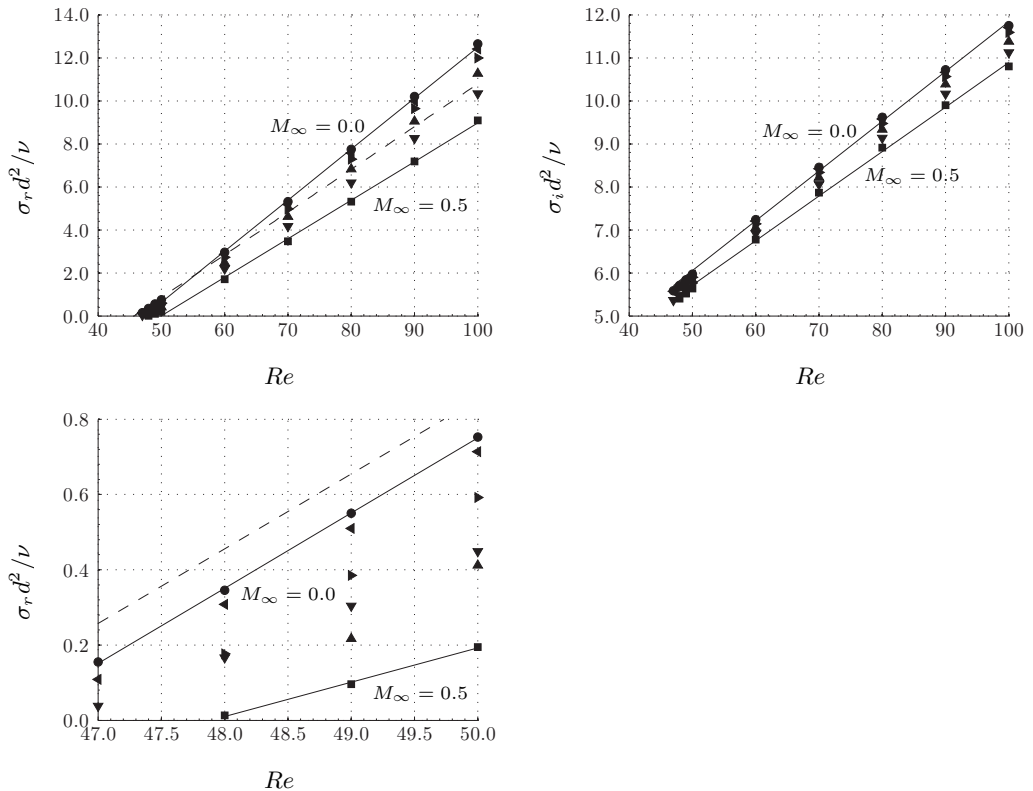
\begin{figure}
  \begin{center}
  \begin{tabular}{m{0.009\textwidth}m{0.48\textwidth}m{0.009\textwidth}m{0.48\textwidth}m{0.0001\textwidth}} 
    \rotatebox[origin=c]{90}{$\sigma_{r}d^2/\nu$}	& 
    {\scriptsize
    \begin{overpic}[width=0.48\textwidth]{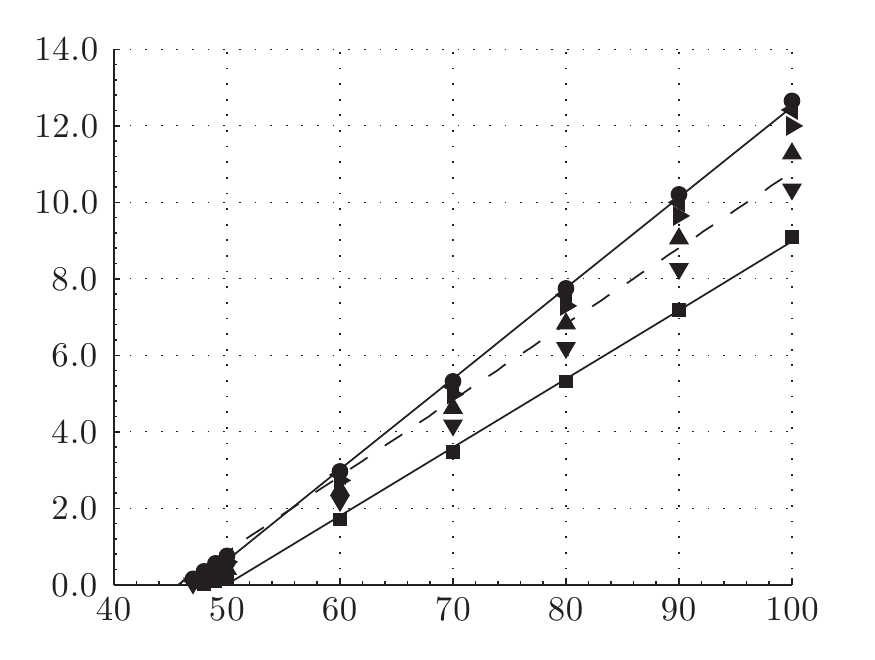}
      \put(48,46){$M_{\infty} = 0.0$}
      \put(69,30){$M_{\infty} = 0.5$}
    \end{overpic}} &
    \rotatebox[origin=c]{90}{$\sigma_{i}d^2/\nu$}	& 
    {\scriptsize
    \begin{overpic}[width=0.48\textwidth]{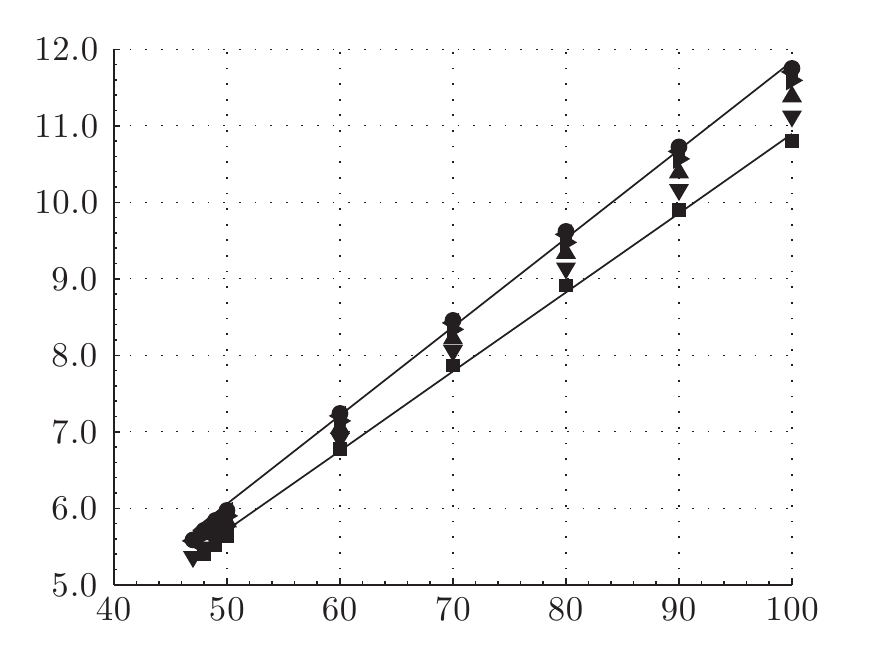}
      \put(40,45){$M_{\infty} = 0.0$}
      \put(61,36){$M_{\infty} = 0.5$}
    \end{overpic}} &  \\
    & \center{$Re$}	& 
    & \center{$Re$} & \\
   \rotatebox[origin=c]{90}{$\sigma_{r}d^2/\nu$}	& 
    {\scriptsize
    \begin{overpic}[width=0.48\textwidth]{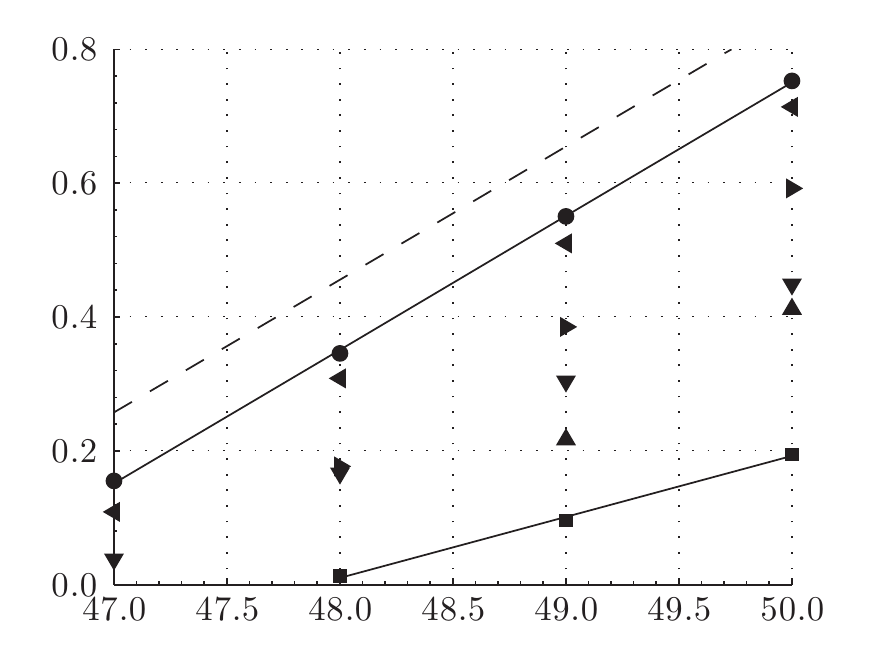}
      \put(42,33){$M_{\infty} = 0.0$}
      \put(69,13){$M_{\infty} = 0.5$}
    \end{overpic}} & & & \\
    & \center{$Re$} 	& & & \\
  \end{tabular}
  \caption{Growth rate $\sigma_{r}d^2/\nu$ (upper left) and dominant unstable frequency $\sigma_{i}d^2/\nu$ (upper right) obtained for various $M_{\infty}$ as functions of $Re$. Overlaid solid lines are linear curve fits of the present data (excluding observations at $Re < 50$), while the dashed line is taken from the incompressible flow data of \cite{Strykowski:JFM90}. At bottom left is an inset of the the growth rate $\sigma_{r}d^2/\nu$ for $Re \le 50$, with selected curve fits excluding observations at $Re > 50$. ({\large{$\bullet$}}) $M_{\infty} = 0.0$; ($\blacktriangleleft$) $M_{\infty} = 0.1$; ($\blacktriangleright$) $M_{\infty} = 0.2$; ($\blacktriangle$) $M_{\infty} = 0.3$; ($\blacktriangledown$) $M_{\infty} = 0.4$; ({\scriptsize{$\blacksquare$}}) $M_{\infty} = 0.5$.}
  \label{fig:sr_linear}
  \end{center}
\end{figure}

For the growth rate of instabilities, we find that compressibility stabilizes the flow, as increasing $M_{\infty}$ decreases $\sigma_r$. We also note that the critical Reynolds number $Re_c$ increases with increasing $M_{\infty}$. These trends can be seen in figure \ref{fig:sr_linear}, which shows $\sigma_{r}d^2/\nu$ as a function of $Re$ for various $M_{\infty}$. \cite{Strykowski:JFM90} made experimental measurements of this quantity in a wind tunnel equipped with valves that allowed for rapid change in the wind tunnel velocity. By changing the velocity, $Re$ was quickly changed from stable to unstable values and vice-versa. Hot-wire anemometry and laser doppler velocimetry recordings of streamwise velocity were then employed in the wake to show that for incompressible flow, $\sigma_{r}d^2/\nu$ can be fitted with a function of the form
\begin{equation}
  \frac{\sigma_{r}d^2}{\nu} = \alpha Re + \beta,
  \label{eqn:sr_linear}
\end{equation}
\noindent where the term $d^2/\nu$ is the viscous timescale for each case. For their analysis, $\alpha$ and $\beta$ were taken to be constants. The present data demonstrates that this correlation works well for $Re$ beyond the predicted $Re_c$, and can also be applied to the non-dimensional dominant unstable frequency 
\begin{equation}
  \frac{\sigma_{i}d^2}{\nu} = \eta Re + \delta,
  \label{eqn:si_linear}
\end{equation}
\noindent where $\eta$ and $\delta$ are constants at a given $M_{\infty}$. However, in both cases, the correlations show slight deviation near $Re_c$. To illustrate this point, we perform curve fits of $\sigma_{r}d^2/\nu$ and $\sigma_{i}d^2/\nu$ as a function of $Re$ at each $M_{\infty}$, excluding observations made at $Re < 50$. Curve fits of the form given in (\ref{eqn:sr_linear}) and (\ref{eqn:si_linear}) are presented for selected cases using solid lines in the upper row of figure \ref{fig:sr_linear}, and show that (\ref{eqn:sr_linear}) and (\ref{eqn:si_linear}) slightly underpredict the growth rate for cases near $Re_c$.

We also observe that the present data for incompressible flow exhibits lower growth rates than those taken from \cite{Strykowski:JFM90} for most of the range of $Re$ considered ($Re \gtrsim 55$). We attribute this discrepancy to differences in the experimental and numerical setups. While the flow in the present study is provided with a small perturbation at constant $Re$, the flow in the study undertaken by \cite{Strykowski:JFM90} had to be quickly altered from stable to unstable $Re$ through a change in the wind tunnel velocity. The transient period resulting from this flow acceleration may explain the difference in growth rates, given that the observed difference increases with increasing $Re$ (i.e., with greater flow acceleration in the wind tunnel). It should be further noted that the initial base state in the analysis performed by \cite{Strykowski:JFM90} is at a stable $Re$, whereas the present study is initialized from a base state with unstable $Re$.

\begin{figure}
  \begin{center}
  \begin{tabular}{m{0.009\textwidth}m{0.48\textwidth}m{0.009\textwidth}m{0.48\textwidth}m{0.0001\textwidth}} 
    \rotatebox[origin=c]{90}{$\alpha$} &
    {\scriptsize
    \begin{overpic}[width=0.48\textwidth]{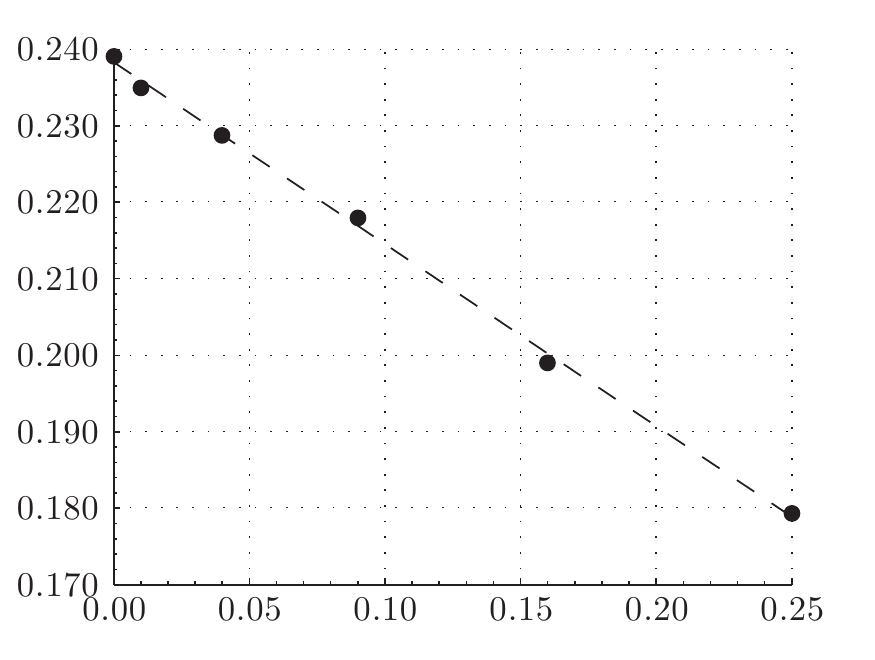} 
      \put(18,45){$\alpha_{0} = 0.237$}
      \put(18,39){$\alpha_{2} = -0.231$}
    \end{overpic}} &
    \rotatebox[origin=c]{90}{$\beta$} & 
    {\scriptsize
    \begin{overpic}[width=0.48\textwidth]{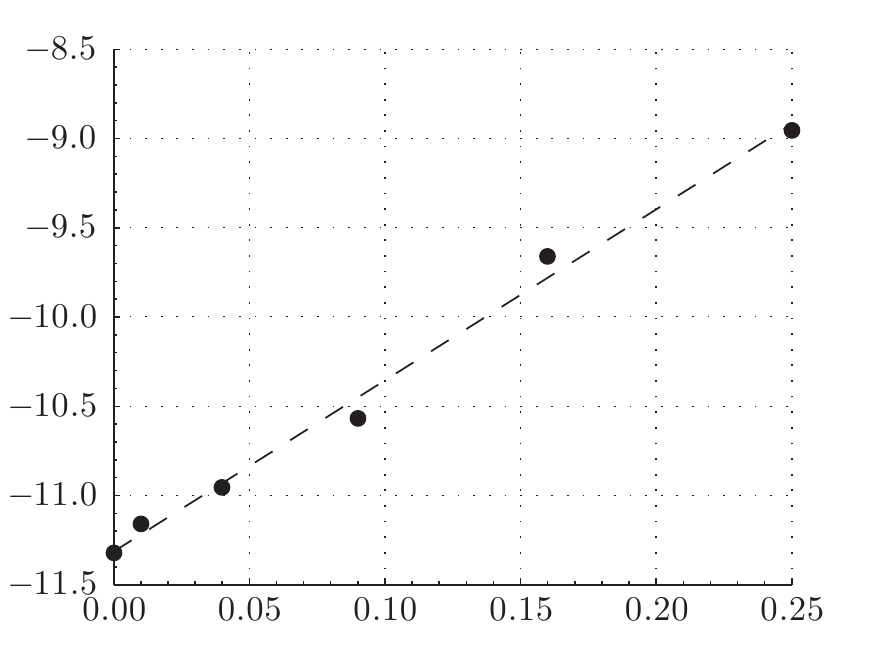}
      \put(21,41){$\beta_{0} = -11.2$}
      \put(21,35){$\beta_{2} = 9.03$}
    \end{overpic}} & \\
    \rotatebox[origin=c]{90}{$\eta$} &
    {\scriptsize
    \begin{overpic}[width=0.48\textwidth]{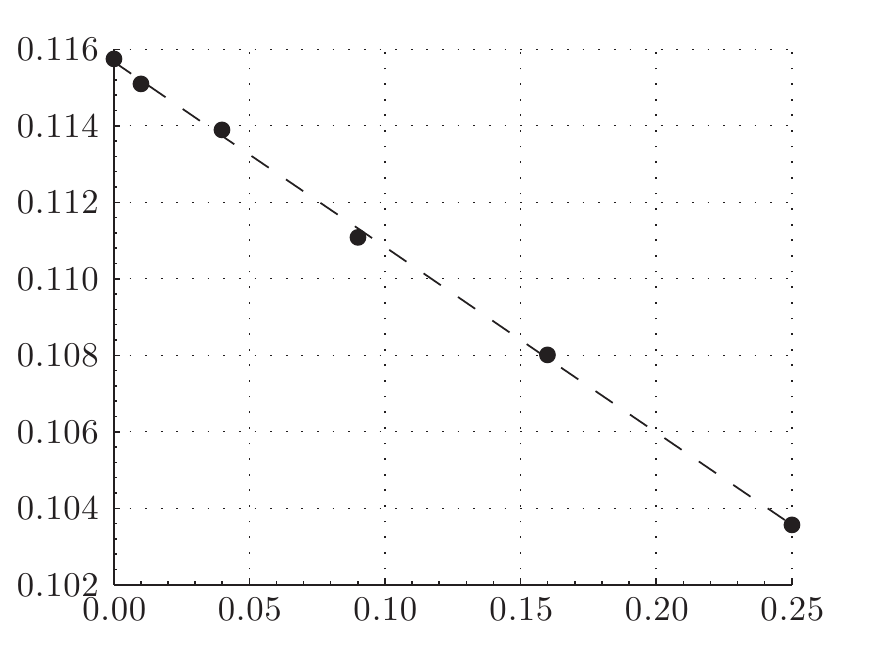} 
      \put(18,44){$\eta_{0} = 0.116$}
      \put(18,38){$\eta_{2} = -0.049$}
    \end{overpic}} &
    \rotatebox[origin=c]{90}{$\delta$} & 
    {\scriptsize
    \begin{overpic}[width=0.48\textwidth]{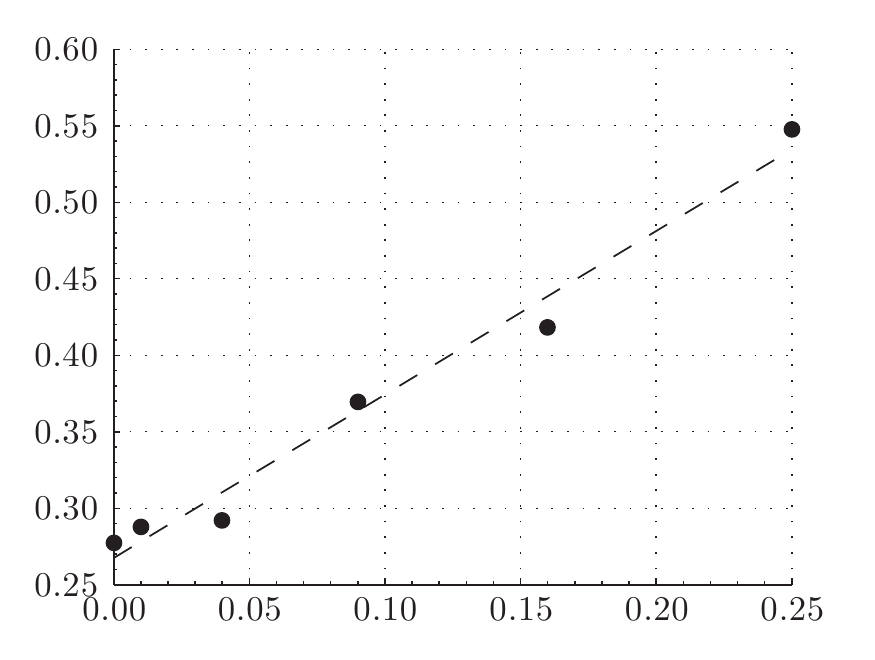}
      \put(21,37){$\delta_{0} = 0.268$}
      \put(21,31){$\delta_{2} = 1.07$}
    \end{overpic}} & \\
    & \center{$M_{\infty}^2$} &							
    & \center{$M_{\infty}^2$} & \\
  \end{tabular}
  \caption{Slopes ($\alpha = \alpha_{2} M_{\infty}^{2} + \alpha_0$, $\eta = \eta_{2} M_{\infty}^{2} + \eta_0$) and intercepts ($\beta = \beta_{2} M_{\infty}^{2} + \beta_0$,   $\delta = \delta_{2} M_{\infty}^{2} + \delta_0$) obtained from linear curve fits of $\sigma_{r}d^2/\nu = \alpha Re + \beta$ and $\sigma_{i}d^2/\nu =\ \eta Re + \delta$.}
  \label{fig:sr_linear_slopes}
  \end{center}
\end{figure}
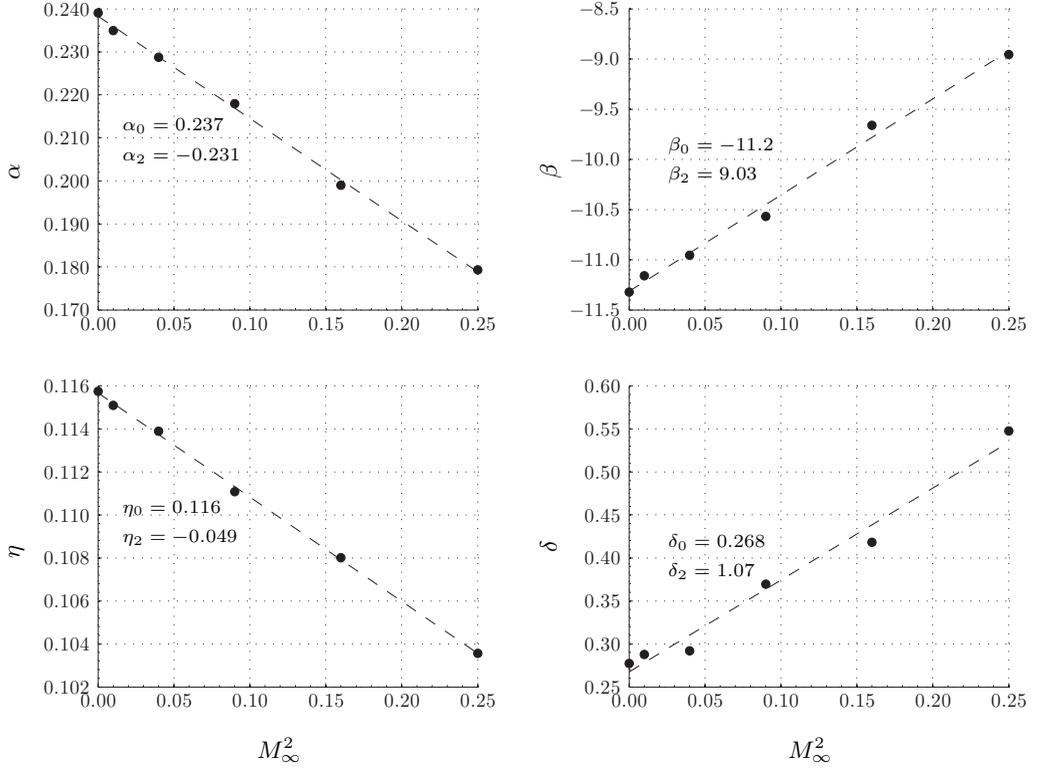

The accuracy of (\ref{eqn:sr_linear}) and (\ref{eqn:si_linear}) over most of the range of $Re$ considered in this study motivates the creation of a similar equation that includes compressibility effects.  Owing to the strong quadratic correlation between $\overline{C}_d$ and $M_{\infty}$ previously observed in figure \ref{fig:Cd_norm}, similar quadratic functions were chosen as first approximations to the dependence of the slopes and intercepts of (\ref{eqn:sr_linear}) and (\ref{eqn:si_linear}) on $M_{\infty}$. Indeed, we find that both the slopes and the intercepts are well-predicted using a function of $M_{\infty}^2$, as shown in figure \ref{fig:sr_linear_slopes}. As such, the constants in (\ref{eqn:sr_linear}) and (\ref{eqn:si_linear}) may be re-written to include compressibility effects:
\begin{equation}
  \alpha = \alpha_{2} M_{\infty}^{2} + \alpha_0,
  \label{eqn:alpha}
\end{equation}
\begin{equation}
  \beta = \beta_{2} M_{\infty}^{2} + \beta_0,
  \label{eqn:beta}
\end{equation}
\begin{equation}
  \eta = \eta_{2} M_{\infty}^{2} + \eta_0,
  \label{eqn:gamma}
\end{equation}
\begin{equation}
  \delta = \delta_{2} M_{\infty}^{2} + \delta_0,
  \label{eqn:delta}
\end{equation}
\noindent where the parameters on the right-hand side are taken to be constant over the range of $M_{\infty}$ considered and are overlaid in figure \ref{fig:sr_linear_slopes}. Substituting these equations into (\ref{eqn:sr_linear}) and (\ref{eqn:si_linear}) allows one to obtain
\begin{equation}
  \frac{\sigma_{r}d^2}{\nu} = (\alpha_{2} M_\infty^2 + \alpha_{0})Re + (\beta_{2} M_\infty^2 + \beta_0),
  \label{eqn:sr_linear_2}
\end{equation}
\begin{equation}
  \frac{\sigma_{i}d^2}{\nu} = (\eta_{2} M_\infty^2 + \eta_{0})Re + (\delta_{2} M_\infty^2 + \delta_0),
  \label{eqn:si_linear_2}
\end{equation}
which enables us to predict the growth rate and dominant frequency of instabilities for two-dimensional compressible laminar flow. Again, it must be noted that these equations have only been validated for $50 \le Re \le 100$ and $M_{\infty} \le 0.5$. We have also considered a separate set of curve fits for observations made at $Re \le 50$ using this same procedure. In this manner, it would have been possible to create piecewise functions capable of predicting the growth rates and frequencies over the entire range of $Re$ and $M_{\infty}$ considered. Indeed, it was found that the data was well-described by equations of the form given in (\ref{eqn:sr_linear}) and (\ref{eqn:si_linear}), and selected curve fits are overlaid in the inset at the bottom-left of figure \ref{fig:sr_linear}. However, as illustrated in the inset, there was a sudden increase in the growth rate from $M_{\infty} = 0.3$ to $0.4$. As such, it appears that a na\"{i}ve curve fit as a function of $M_{\infty}^2$, as in (\ref{eqn:alpha}) through (\ref{eqn:delta}), would not fully capture the transition.

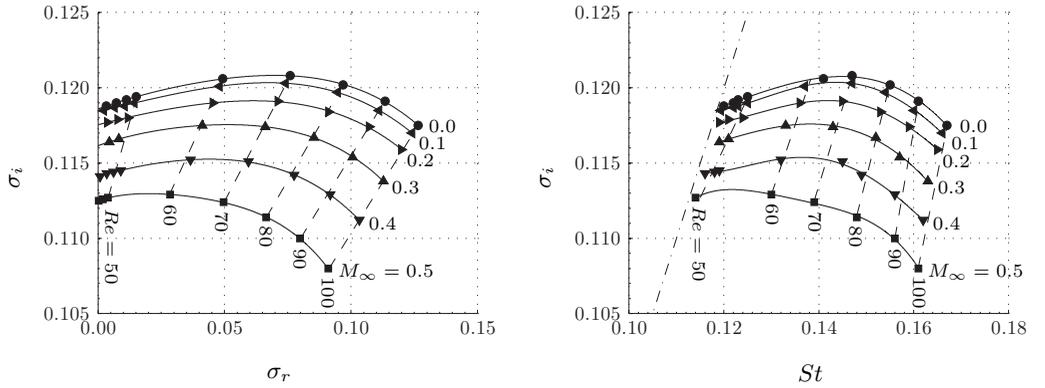
\begin{figure}
  \begin{center}
  \begin{tabular}{m{0.009\textwidth}m{0.48\textwidth}m{0.009\textwidth}m{0.48\textwidth}m{0.0001\textwidth}} 
    \rotatebox[origin=c]{90}{$\sigma_{i}$} &
    {\scriptsize
    \begin{overpic}[width=0.48\textwidth]{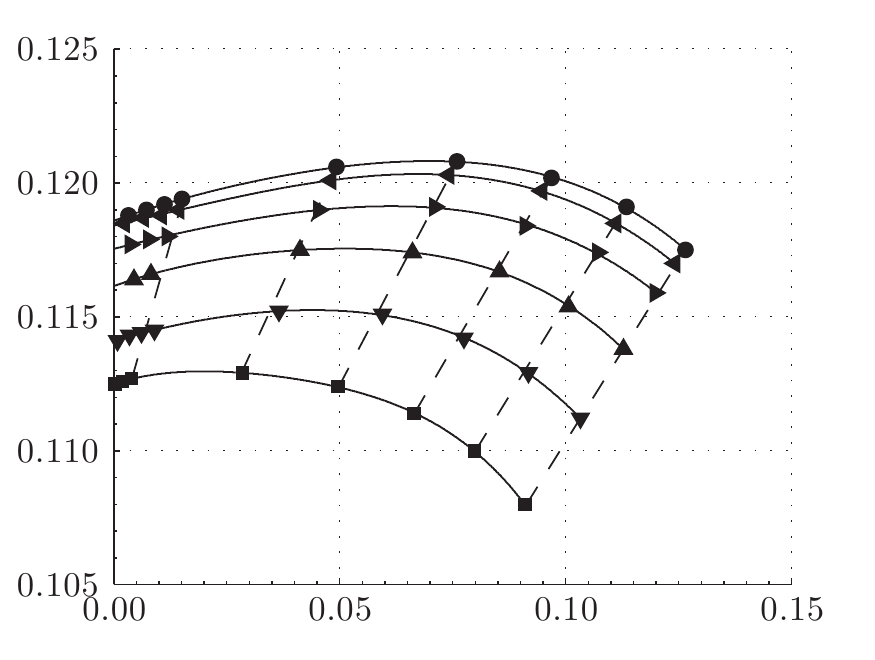}
      \put(14.5,21){\rotatebox[origin=c]{-90}{$Re = 50$}}
      \put(26.5,27.3){\rotatebox[origin=c]{-90}{$60$}}
      \put(37,26){\rotatebox[origin=c]{-90}{$70$}}
      \put(46,23){\rotatebox[origin=c]{-90}{$80$}}
      \put(53,18.5){\rotatebox[origin=c]{-90}{$90$}}
      \put(58.7,11.7){\rotatebox[origin=c]{-90}{$100$}}
      \put(62,16){$M_{\infty} = 0.5$}
      \put(68.3,25.7){$0.4$}
      \put(73,33){$0.3$}
      \put(76,38.5){$0.2$}
      \put(78.5,42){$0.1$}
      \put(80.5,45.5){$0.0$}
    \end{overpic}} &
    \rotatebox[origin=c]{90}{$\sigma_{i}$} &
    {\scriptsize
    \begin{overpic}[width=0.48\textwidth]{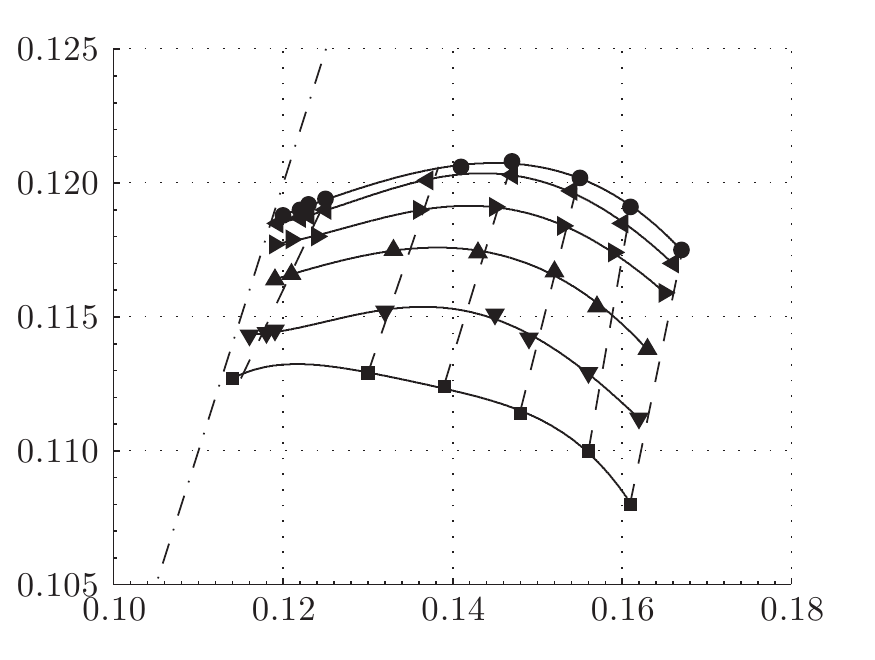}
      \put(26,21.5){\rotatebox[origin=c]{-90}{$Re = 50$}}
      \put(41,27.6){\rotatebox[origin=c]{-90}{$60$}}
      \put(49.7,26){\rotatebox[origin=c]{-90}{$70$}}
      \put(58,23){\rotatebox[origin=c]{-90}{$80$}}
      \put(66,18.5){\rotatebox[origin=c]{-90}{$90$}}
      \put(71,11.7){\rotatebox[origin=c]{-90}{$100$}}
      \put(74,16){$M_{\infty}=0.5$}
      \put(75,25.7){$0.4$}
      \put(76,33){$0.3$}
      \put(77,38.3){$0.2$}
      \put(78.8,42){$0.1$}
      \put(80.5,45.5){$0.0$}
    \end{overpic}} & \\
    & \center{$\sigma_r$} & 
    & \center{$St$} & \\
  \end{tabular}
  \caption{Dominant unstable frequency $\sigma_i$ obtained for various $M_{\infty}$ as functions of $\sigma_{r}$ (left). Comparison of $\sigma_i$ and $St$ for various $M_{\infty}$ (right).  Reference line (dashed-dotted) corresponds to $\sigma_{i} = St$.}
  \label{fig:sigma_polar}
  \end{center}
\end{figure}

The impact of compressibility on the frequency of oscillations during the linear growth stage (measured by $\sigma_i$) can be observed in figure \ref{fig:sigma_polar}.  From this plot, we see that as $Re$ increases, the frequency of oscillations increases from the linear growth stage to the saturated nonlinear oscillatory flow. This result is not unexpected, since it is well-established that the frequency of instabilities increases in conjunction with $Re$ for the considered range. However, for fixed $Re$ and increasing $M_{\infty}$, we also observe that the frequency of oscillations in the linear growth stage decreases. This trend matches that previously discussed for flows in the nonlinear, saturated stage (figure \ref{fig:reducedSt_M}). Furthermore, over the range of $M_{\infty}$ considered in this study, we find that the change in $\sigma_i$ relative to its incompressible value remains relatively constant with increasing $Re$. By contrast, the relative change in $St$ over the considered range of $M_{\infty}$ becomes smaller as $Re$ increases, as previously discussed in the context of figure \ref{fig:reducedSt_M}.

\begin{figure}
  \begin{center}
  \begin{tabular}{m{0.009\textwidth}m{0.48\textwidth}m{0.0001\textwidth}} 
    \rotatebox[origin=c]{90}{$\sigma_{i,c}$} & 
    {\scriptsize
    \begin{overpic}[width=0.48\textwidth]{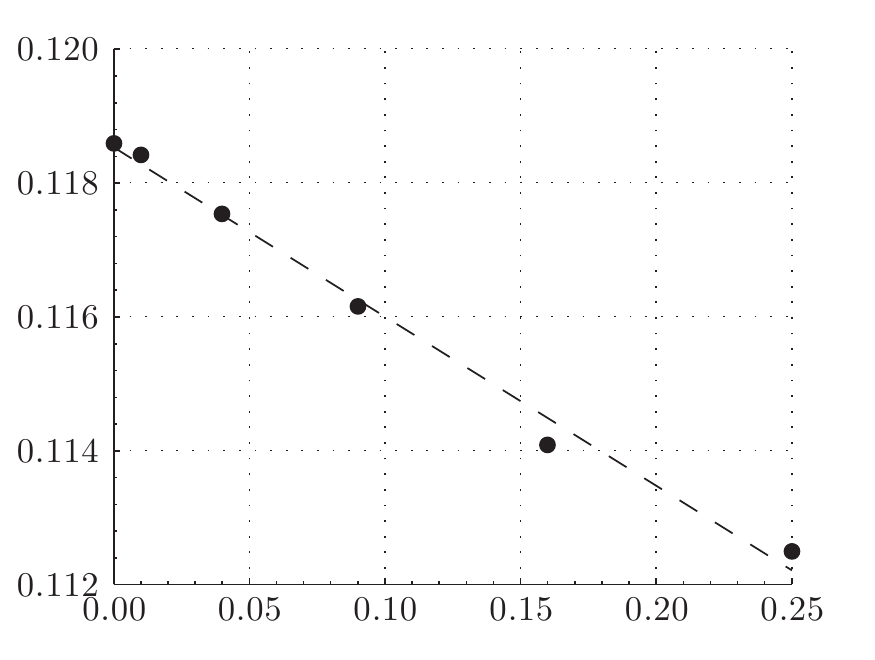}
    \end{overpic}} & \\
    & \center{$M_{\infty}^2$} & \\
  \end{tabular}
  \caption{Critical dominant frequency $\sigma_{i,c} = \sigma_i\rvert_{\sigma_r = 0}$ as a function of $M_{\infty}^2$. }
  \label{fig:sigma_ic}
  \end{center}
\end{figure}

Interestingly, the critical dominant frequencies $\sigma_{i,c}$ (i.e., $\sigma_i$ when $\sigma_r = 0$) predicted by the trendlines given on the left-hand side of figure \ref{fig:sigma_polar} seem to exhibit a quadratic relationship with $M_{\infty}$, as displayed in figure \ref{fig:sigma_ic}. This relationship holds despite the fact that the trendlines take into account growth rates for which $(Re - Re_c) < 5$, a condition that we observed to result in deviations from the linear correlation between $\sigma_{r}d^2/\nu$ and $Re$ presented in (\ref{eqn:sr_linear}).

At last, we note that the data used in this work is provided in Appendix A.  Tables \ref{tab:kd} and \ref{tab:kd2} summarize the wake characteristics behind a circular cylinder for compressible viscous flow at representative Reynolds numbers of 20 to 60 and 70 to 100, respectively.  Table \ref{tab:growth} lists the linear growth rates and frequencies for instabilities with respect to the compressible unstable steady states around a circular cylinder.

%%%%%%%%%%%%%%%%%%%%%%%%%%%%%%%%%%%%%%%%%%%%%%%%%%%%%%%%%%%%%%%%%%

\section{Conclusion}
\label{sec:conclusion}

The present study employs direct numerical simulation to examine the two-dimensional, viscous, compressible flow around a circular cylinder. In comparing flowfield characteristics, interesting similarities are shown between stable and unstable cases. In both regimes, it is found that increasing the freestream Mach number increases $\overline{C}_d$. This result is expected, and reflects results from other studies at higher $Re$ \citep{Lindsey:NACA619, Macha:JOA77}. For cases below the critical Reynolds number for two-dimensional shedding, it is found that compressibility effects elongate the wake and cause a small delay in flow separation. Likewise, for unstable cases, increasing the freestream Mach number results in a larger time-averaged recirculation region and vortex shedding wavelength. Furthermore, as $Re$ is decreased for stable flows, the wake geometry is observed to exhibit a smaller variation over the range of freestream Mach numbers considered. In similar fashion for unstable flows, though the non-dimensional shedding frequency decreases monotonically with increasing freestream Mach number at every $Re$ studied, the magnitude of this decrease shrinks as $Re$ increases. These similarities in behavior between stable and unstable flows suggest that the difference between $Re$ and the critical Reynolds number for the emergence of two-dimensional shedding plays a significant role in determining the sensitivity of a flow to compressibility effects.

Through linear stability analysis, the growth rate and frequency of the primary wake instability that arises for $Re$ above the critical Reynolds number are examined. In doing so, it is found that the non-dimensional growth rate of the primary wake instability exhibits a linear correlation with $Re$ at every freestream Mach number considered. Moreover, the variation in the curve fitting parameters for correlations at different freestream Mach numbers are found to be well-explained by a dependence on the square of the Mach number. Additionally, it is observed that at a given $Re$, the growth rate decreases with increasing freestream Mach number. This trend is also observed for the dominant unstable frequency and the non-dimensional shedding frequency, hence indicating that compressibility effects tend to stabilize the flow. Finally, it is seen that as $Re$ approaches the critical Reynolds number, the non-dimensional shedding frequency approaches the dominant unstable frequency. This behavior suggests that as $Re$ nears the critical Reynolds number, the saturated flow characteristics increasingly depend on the characteristics of the dominant unstable mode in the linear growth stage. This result, coupled with the previous observation that the flow also becomes more sensitive to compressibility effects as $Re$ approaches the critical value, suggests that the dominant unstable mode is sensitive to changes in the freestream Mach number.

%%%%%%%%%%%%%%%%%%%%%%%%%%%%%%%%%%%%%%%%%%%%%%%%%%%%%%%%%%%%%%%%%%

\section*{Acknowledgements}

DC acknowledges partial support from the National Science Foundation (NSF-REU Program, Award Number: 1062936). DC and KT also acknowledge the support from the US Army Research Office (Grant Number W911NF-13-1-0146).  Finally, the authors would like to thank Phillip Munday for his technical guidance.

%%%%%%%%%%%%%%%%%%%%%%%%%%%%%%%%%%%%%%%%%%%%%%%%%%%%%%%%%%%%%%%%%%

\appendix
\section{}\label{appA}
This appendix contains the data presented in this paper tabulated as Tables \ref{tab:kd}, \ref{tab:kd2}, and \ref{tab:growth}.

\begin{table}
  \begin{center}
\def~{\hphantom{0}}
  \begin{tabular}{ccccccccc}
      $Re$  	& $M_\infty$ 	& $C_d$ 				& $C_l$ 			& $l_a/d$ 	& $l_b/d$ 	& $l_r/d$	& $\theta_{s}$ 	& $St$ 	\\ \hline
       	20	& 0  		  	& 2.07 				& -				& 0.356	& 0.425	& 0.921	& 43.7$^\circ$	& -		\\
       		& 0.1 			& 2.09 				& -				& 0.356	& 0.425	& 0.922	& 43.7$^\circ$	& -		\\
		& 0.2 			& 2.08 				& -				& 0.358	& 0.426	& 0.928	& 43.7$^\circ$	& -		\\
		& 0.3 			& 2.11 				& -				& 0.362	& 0.428	& 0.940	& 43.7$^\circ$	& -		\\
		& 0.4 			& 2.16 				& -				& 0.365	& 0.429	& 0.954	& 43.6$^\circ$	& -		\\
		& 0.5 			& 2.23 				& -				& 0.367	& 0.428	& 0.963	& 43.4$^\circ$	& -		\\ \hline
		
       	40	& 0  		  	& 1.54 				& -				& 0.715	& 0.594	& 2.24	& 53.8$^\circ$	& -		\\
       		& 0.1 			& 1.54 				& -				& 0.718	& 0.594	& 2.25	& 53.7$^\circ$	& -		\\
		& 0.2 			& 1.56 				& -				& 0.724	& 0.596	& 2.28	& 53.7$^\circ$	& -		\\
		& 0.3 			& 1.58 				& -				& 0.736	& 0.602	& 2.34	& 53.7$^\circ$	& -		\\
		& 0.4 			& 1.62				& -				& 0.752	& 0.609	& 2.43	& 53.5$^\circ$	& -		\\
		& 0.5 			& 1.68 				& -				& 0.769	& 0.614	& 2.53	& 53.3$^\circ$	& -		\\ \hline	
		
       	50	& 0  		& 1.44 $\pm$ 0.00160 			& $\pm$ 0.0561	& - 		& -		& -		& -			& 0.126	\\
       		& 0.1 		& 1.45 $\pm$ 0.00155 			& $\pm$ 0.0546	& - 		& -		& -		& -			& 0.125	\\
		& 0.2 		& 1.46 $\pm$ 0.00122 			& $\pm$ 0.0486	& - 		& -		& -		& -			& 0.124	\\
		& 0.3 		& 1.46 $\pm$ 0.00115 			& $\pm$ 0.0399	& - 		& -		& -		& -			& 0.121	\\
		& 0.4 		& 1.52 $\pm$ 4.53 $\times 10^{-5}$  	& $\pm$ 0.0314	& - 		& -		& -		& -			& 0.119	\\
		& 0.5 		& 1.56 $\pm$ 2.03 $\times 10^{-5}$  	& $\pm$ 0.0144	& - 		& -		& -		& -			& 0.114	\\ \hline
		
	60	& 0  		& 1.41 $\pm$ 0.00105 			& $\pm$ 0.132		& - 		& -		& -		& -			& 0.141	\\
       		& 0.1 		& 1.416 $\pm$ 0.00105 			& $\pm$ 0.132		& - 		& -		& -		& -			& 0.137	\\
		& 0.2 		& 1.427 $\pm$ 0.00095			& $\pm$ 0.125		& - 		& -		& -		& -			& 0.136	\\
		& 0.3 		& 1.453 $\pm$ 0.0009			& $\pm$ 0.118		& - 		& -		& -		& -			& 0.133	\\
		& 0.4 		& 1.4925 $\pm$ 0.00065		  	& $\pm$ 0.112		& - 		& -		& -		& -			& 0.132	\\
		& 0.5 		& 1.5510 $\pm$ 0.0005  			& $\pm$ 0.091		& - 		& -		& -		& -			& 0.130	\\ \hline
     \end{tabular}
  \caption{Wake characteristics behind a circular cylinder for compressible viscous flow at representative Reynolds numbers (20 to 60).}
  \label{tab:kd}
  \end{center}
\end{table}

\begin{table}
  \begin{center}
\def~{\hphantom{0}}
  \begin{tabular}{ccccccccc}
      $Re$  	& $M_\infty$ 	& $C_d$ 						& $C_l$ 			& $l_a/d$ 	& $l_b/d$ 	& $l_r/d$	& $\theta_{s}$ 	& $St$ 	\\ \hline
      
	70	& 0  			& 1.383 $\pm$ 0.0023 			& $\pm$ 0.190		& - 		& -		& -		& -			& 0.147	\\
       		& 0.1 		& 1.392 $\pm$ 0.0025			& $\pm$ 0.194		& - 		& -		& -		& -			& 0.147	\\
		& 0.2 		& 1.403 $\pm$ 0.0024			& $\pm$ 0.186		& - 		& -		& -		& -			& 0.145	\\
		& 0.3 		& 1.430 $\pm$ 0.0022			& $\pm$ 0.178		& - 		& -		& -		& -			& 0.143	\\
		& 0.4 		& 1.503 $\pm$ 0.0056		  	& $\pm$ 0.230		& - 		& -		& -		& -			& 0.145	\\
		& 0.5 		& 1.535 $\pm$ 0.0013  			& $\pm$ 0.156		& - 		& -		& -		& -			& 0.139	\\ \hline

      	75	& 0  		  	& 1.37 $\pm$ 0.00313 			& $\pm$ 0.216		& -	 	& -		& -		& -			& 0.151	\\
       		& 0.1 		& 1.38 $\pm$ 0.00336 			& $\pm$ 0.217		& -	 	& -		& -		& -			& 0.150	\\
		& 0.2 		& 1.39 $\pm$ 0.00310 			& $\pm$ 0.212		& -	 	& -		& -		& -			& 0.150	\\
		& 0.3 		& 1.42 $\pm$ 0.00288 			& $\pm$ 0.205		& -	 	& -		& -		& -			& 0.148	\\
		& 0.4 		& 1.47 $\pm$ 0.00356 			& $\pm$ 0.210		& -	 	& -		& -		& -			& 0.147	\\
		& 0.5 		& 1.53 $\pm$ 0.00214 			& $\pm$ 0.190		& -	 	& -		& -		& -			& 0.144	\\ \hline
      
      	80	& 0  			& 1.364 $\pm$ 0.0040 			& $\pm$ 0.240		& - 		& -		& -		& -			& 0.155	\\
       		& 0.1 		& 1.373 $\pm$ 0.0040			& $\pm$ 0.239		& - 		& -		& -		& -			& 0.154	\\
		& 0.2 		& 1.385 $\pm$ 0.0041			& $\pm$ 0.238		& - 		& -		& -		& -			& 0.153	\\
		& 0.3 		& 1.413 $\pm$ 0.0040			& $\pm$ 0.234		& - 		& -		& -		& -			& 0.152	\\
		& 0.4 		& 1.466 $\pm$ 0.0045		  	& $\pm$ 0.230		& - 		& -		& -		& -			& 0.149	\\
		& 0.5 		& 1.525 $\pm$ 0.0024  			& $\pm$ 0.222		& - 		& -		& -		& -			& 0.148	\\ \hline
            		
	90	& 0  			& 1.350 $\pm$ 0.0065 			& $\pm$ 0.286		& - 		& -		& -		& -			& 0.161	\\
       		& 0.1 		& 1.359 $\pm$ 0.0065			& $\pm$ 0.286		& - 		& -		& -		& -			& 0.160	\\
		& 0.2 		& 1.373 $\pm$ 0.0059			& $\pm$ 0.286		& - 		& -		& -		& -			& 0.159	\\
		& 0.3 		& 1.404 $\pm$ 0.0066			& $\pm$ 0.286		& - 		& -		& -		& -			& 0.157	\\
		& 0.4 		& 1.452 $\pm$ 0.0056		  	& $\pm$ 0.284		& - 		& -		& -		& -			& 0.156	\\
		& 0.5 		& 1.541 $\pm$ 0.0090  			& $\pm$ 0.279		& - 		& -		& -		& -			& 0.156	\\ \hline
		
       100	& 0 			& 1.34 $\pm$ 0.0091 			& $\pm$ 0.329		& -	 	& -		& -		& -			& 0.167  	\\
       		& 0.1 		& 1.35 $\pm$ 0.0096 			& $\pm$ 0.328		& -	 	& -		& -		& -			& 0.166  	\\
		& 0.2 		& 1.36 $\pm$ 0.0093 			& $\pm$ 0.332		& -	 	& -		& -		& -			& 0.165  	\\
		& 0.3 		& 1.39 $\pm$ 0.0086 			& $\pm$ 0.326		& -	 	& -		& -		& -			& 0.163  	\\
		& 0.4 		& 1.45 $\pm$ 0.0086 			& $\pm$ 0.328		& -	 	& -		& -		& -			& 0.162  	\\
		& 0.5 		& 1.54 $\pm$ 0.0101 			& $\pm$ 0.352		& -	 	& -		& -		& -			& 0.161  	\\ \hline
  \end{tabular}
  \caption{Wake characteristics behind a circular cylinder for compressible viscous flow at representative Reynolds numbers (70 to 100).}
  \label{tab:kd2}
  \end{center}
\end{table}

\begin{table}
	\begin{center}
		\def~{\hphantom{0}}
		\begin{tabular}{ccc}
			\begin{tabular}{cccc}
				$Re$	& $M_\infty$	& $\sigma_{r}$		& $\sigma_{i}$	\\ \hline
				47		& 0			& 3.037 $\times 10^{-3}$ 	& 0.1188			\\
				47		& 0.1			& 2.319 $\times 10^{-3}$ 	& 0.1185			\\
				47		& 0.4			& 7.963 $\times 10^{-4}$ 	& 0.1141			\\ \hline
				48		& 0			& 7.196 $\times 10^{-3}$ 	& 0.1190			\\
				48		& 0.1			& 6.418 $\times 10^{-3}$ 	& 0.1187			\\
				48		& 0.2			& 3.681 $\times 10^{-3}$ 	& 0.1177			\\
				48		& 0.4			& 3.456 $\times 10^{-3}$ 	& 0.1143			\\
				48		& 0.5			& 2.662 $\times 10^{-4}$ 	& 0.1125			\\ \hline
				49		& 0			& 1.123 $\times 10^{-2}$ 	& 0.1192			\\
				49		& 0.1			& 1.041 $\times 10^{-2}$ 	& 0.1188			\\
				49		& 0.2			& 7.854 $\times 10^{-3}$ 	& 0.1179			\\
				49		& 0.3			& 4.436 $\times 10^{-3}$ 	& 0.1164			\\
				49		& 0.4			& 6.193 $\times 10^{-3}$ 	& 0.1144			\\
				49		& 0.5			& 1.964 $\times 10^{-3}$ 	& 0.1126			\\ \hline
				50		& 0			& 1.505 $\times 10^{-2}$ 	& 0.1194			\\ 
				50		& 0.1			& 1.428 $\times 10^{-2}$ 	& 0.1190			\\
				50		& 0.2			& 1.184 $\times 10^{-2}$	& 0.1180			\\
				50		& 0.3			& 8.232 $\times 10^{-3}$	& 0.1166			\\
				50		& 0.4			& 8.974 $\times 10^{-3}$	& 0.1145			\\
				50		& 0.5			& 3.894 $\times 10^{-3}$	& 0.1127			\\ \hline
				60		& 0			& 4.931 $\times 10^{-2}$	& 0.1206			\\
				60		& 0.1			& 4.790 $\times 10^{-2}$	& 0.1201			\\
				60		& 0.2			& 4.536 $\times 10^{-2}$	& 0.1190			\\
				60		& 0.3			& 4.123 $\times 10^{-2}$	& 0.1175			\\
				60		& 0.4			& 3.651 $\times 10^{-2}$	& 0.1152			\\
				60		& 0.5			& 2.844 $\times 10^{-2}$	& 0.1129			\\ \hline
			\end{tabular}
			& ~~~ &
			\begin{tabular}{cccc}
				$Re$	& $M_\infty$	& $\sigma_{r}$		& $\sigma_{i}$	\\ \hline
				70		& 0			& 7.595 $\times 10^{-2}$	& 0.1208			\\
				70		& 0.1			& 7.409 $\times 10^{-2}$	& 0.1203			\\
				70		& 0.2			& 7.102 $\times 10^{-2}$	& 0.1191			\\
				70		& 0.3			& 6.610 $\times 10^{-2}$	& 0.1174			\\
				70		& 0.4			& 5.941 $\times 10^{-2}$	& 0.1151			\\
				70		& 0.5			& 4.964 $\times 10^{-2}$	& 0.1124			\\ \hline
				80		& 0			& 9.686 $\times 10^{-2}$	& 0.1202			\\
				80		& 0.1			& 9.469 $\times 10^{-2}$	& 0.1197			\\
				80		& 0.2			& 9.111 $\times 10^{-2}$	& 0.1184			\\
				80		& 0.3			& 8.536 $\times 10^{-2}$	& 0.1167			\\
				80		& 0.4			& 7.744 $\times 10^{-2}$	& 0.1142			\\
				80		& 0.5			& 6.640 $\times 10^{-2}$	& 0.1114			\\
				90		& 0			& 1.134 $\times 10^{-1}$	& 0.1191			\\ \hline
				90		& 0.1			& 1.110 $\times 10^{-1}$	& 0.1185			\\
				90		& 0.2			& 1.071 $\times 10^{-1}$	& 0.1174			\\
				90		& 0.3			& 1.006 $\times 10^{-1}$	& 0.1154			\\
				90		& 0.4			& 9.174 $\times 10^{-2}$	& 0.1129			\\
				90		& 0.5			& 7.976 $\times 10^{-2}$	& 0.1100			\\ \hline
				100		& 0			& 1.265 $\times 10^{-1}$	& 0.1175			\\
				100		& 0.1			& 1.241 $\times 10^{-1}$	& 0.1170			\\
				100		& 0.2			& 1.199 $\times 10^{-1}$	& 0.1159			\\
				100		& 0.3			& 1.128 $\times 10^{-1}$	& 0.1138			\\
				100		& 0.4			& 1.033 $\times 10^{-1}$	& 0.1112			\\
				100		& 0.5			& 9.095 $\times 10^{-2}$	& 0.1080			\\ \hline
			\end{tabular}
		\end{tabular}
		\caption{Linear growth rates and frequencies for instabilities with respect to compressible unstable steady states around a circular cylinder.}
		\label{tab:growth}
	\end{center}
\end{table}

%%%%%%%%%%%%%%%%%%%%%%%%%%%%%%%%%%%%%%%%%%%%%%%%%%%%%%%%%%%%%%%%%%

\bibliographystyle{jfm}
\bibliography{Taira_all}

\end{document}